\documentclass[aip,pof,english,preprint,preprintnumbers]{revtex4-1}
\usepackage[T1]{fontenc}
\usepackage[latin9]{inputenc}
\usepackage{array}
\usepackage{booktabs}
\usepackage{graphicx}
\usepackage{nomencl}

\makeatletter

\providecommand{\tabularnewline}{\\}

\@ifundefined{textcolor}{}
{%
 \definecolor{BLACK}{gray}{0}
 \definecolor{WHITE}{gray}{1}
 \definecolor{RED}{rgb}{1,0,0}
 \definecolor{GREEN}{rgb}{0,1,0}
 \definecolor{BLUE}{rgb}{0,0,1}
 \definecolor{CYAN}{cmyk}{1,0,0,0}
 \definecolor{MAGENTA}{cmyk}{0,1,0,0}
 \definecolor{YELLOW}{cmyk}{0,0,1,0}
}

\usepackage{siunitx}

\usepackage{url}

\@ifundefined{showcaptionsetup}{}{%
 \PassOptionsToPackage{caption=false}{subfig}}
\usepackage{subfig}
\makeatother

\usepackage{babel}

\usepackage{placeins}

\begin{document}

\preprint{}


\title{The influence of surfactants on thermocapillary flow instabilities in low Prandtl melting pools}

\author{Anton \surname{Kidess}}

\email[To whom correspondence should be addressed: ]{A.Kidess@tudelft.nl}

\author{Sa\v{s}a Kenjere\v{s}}

\author{Chris R. Kleijn}

\affiliation{Delft University of Technology, Department of Chemical Engineering, Transport Phenomena Group}

\date{\today}
\begin{abstract}
Flows in low Prandtl number liquid pools are relevant for various technical applications, and have so far only been investigated for the case of pure fluids, i.e. with a constant, negative surface tension temperature coefficient $\partial\gamma/\partial T$. Real-world fluids containing surfactants have a temperature dependent $\partial\gamma/\partial T > 0$, which may change sign to $\partial\gamma/\partial T < 0$ at a critical temperature $T_c$. Where thermocapillary forces are the main driving force, this can have a tremendous effect on the resulting flow patterns and the associated heat transfer. 

Here we investigate the stability of such flows for five Marangoni numbers in the range of $\num{2.1e6} \leq Ma \leq \num{3.4e7}$ using dynamic large eddy simulations (LES), which we validate against a high resolution direct numerical simulation (DNS). We find that the five cases span all flow regimes, i.e. stable laminar flow at $Ma \leq \num{2.1e6}$, transitional flow with rotational instabilities at $Ma = \num{2.8e6}$ and $Ma = \num{4.6e6}$ and turbulent flow at $Ma = \num{1.8e7}$ and $Ma = \num{3.4e7}$.

This article has been published in Phys. Fluids 28, 062106 (2016), doi:10.1063/1.4953797.
\end{abstract}



\maketitle

\section{Introduction\label{sec:Section-I}}

Flows in heated liquid pools with a free surface are of interest in various technical applications, such as metals processing including fusion welding, electron beam evaporation, casting and crystal growth. In these processes, low Prandtl number liquids are subjected to large surface temperature gradients, which, through resulting gradients in surface tension $\gamma$, lead to thermocapillary motion of the liquid.
In spite of the relevance of low Prandtl number liquids, most research has been devoted to high Prandtl number liquids such as silicone oils \citep{Schatz2001EXPERIMENTS}, as these are easy to work with experimentally and resistant to contamination which can have a major effect on the flow.
Where instabilities in low Prandtl number liquids have been investigated, it has been done for pure materials, i.e. with a constant, negative surface tension temperature coefficient $\partial\gamma/\partial T$, in various geometric configurations (e.g. rectangular\citep{Ohnishi1992Computer,Morvan1996Oscillatory,Bucchignani2004RayleighMarangoni}, liquid bridges \citep{Levenstam2001Instabilities,Lappa2003Threedimensional,Kamotani2007Recent}, annular\citep{Li2004Threedimensional} or  cylindrical\citep{Xu2007RayleighBenardMarangoni}).


However, real-world applications often involve (i) non-pure materials, i.e. materials containing surface active elements (surfactants), and (ii) melting/solidification phase change, both of which influence the thermocapillary flow. The presence of surfactants, even when homogeneously distributed, can have a tremendous effect on the thermocapillary motion by introducing a temperature dependence of the surface tension temperature coefficient $\partial\gamma / \partial T$. Even traces of surfactants such as sulfur or oxygen can be sufficient to introduce a sign change from positive to negative $\partial\gamma / \partial T$ at a critical temperature $T_c$ in many materials\citep{Sahoo1988Surface,Hibiya2010Oxygen}, such as iron, silver, copper or nickel.

The stability of thermocapillary flows in such non-pure low Prandtl number liquids has not been thoroughly studied. \citet{Azami2001Effect} used X-ray tomography to probe the flow of liquid silicon in a half-zone liquid bridge subject to a controlled atmosphere containing varying concentrations of oxygen, which is surface active in silicon and many other liquid metals. They found the flow to be chaotic for low oxygen partial pressures, and to stabilize when the partial pressure is increased above a certain threshold. Here, the oxygen, after adsorption to the silicon, alters the value of $\partial\gamma / \partial T$, which however remains negative and may be regarded as constant, as the applied temperature difference is fairly low. Zhao et al., using particle image velocimetry (PIV) to image the movement of oxide particles floating on the free surface, examined the thermocapillary flow in liquid steel melt pool, heated by a translating electric arc \citep{Zhao2009Unsteady} or a stationary laser \citep{Zhao2010Effect}. For the translating arc, where the surfactant was present as an oxide layer on the surface, they observed an elliptic pool shape subject to periodic non-symmetric oscillations of the surface flow. For stationary laser heating, where oxygen was introduced through a controlled atmosphere, they observed a circular melt pool in which the surface flow was asymmetric and rotational. \citet{kou2011oscillatory}, using qualitative video analysis of the surface of steel melts with sulfur impurities, observe free surface oscillations, indicating instationary flow phenomena. Previous numerical studies of thermocapillary flows taking into account the presence of surfactants and phase change, e.g. \citet{Winker2005Multicomponent} and \citet{DoQuang2008Modeling}, fail to report instabilities observed in the aforementioned experiments, probably due to imposed symmetry and the use of diffusive numerical schemes.

The available literature on the instability of low Prandtl number thermocapillary flows with surfactants is thus exclusively experimental, and rather qualitative. We aim to improve the understanding of instabilities arising in such flows by providing insight into the flow in the pool, addressing the possible occurrence of flow instabilities and turbulence. Specifically, we will investigate thermocapillary flows with non-uniform axisymmetric heating at the free surface such that a negative radial temperature gradient $dT/dr$ develops across the free surface. 

For pure low Prandtl fluids with constant, negative $\partial\gamma / \partial T$ in such cylindrical domains, the onset of oscillatory and turbulent convection is already well understood. \citet{Pumir1996Heat} derived scaling laws for the dimensionless heat transport assuming laminar ($Nu\sim Ma^{1/4}Pr^{1/2}$) or turbulent ($Nu\sim Ma^{1/3}Pr^{1/3}$) thermocapillary flow. Here, the Marangoni number is defined as $Ma=\partial\gamma / \partial T P (\rho\nu\alpha\lambda)^{-1}$, with $P$, $\rho$, $\nu$, $\alpha$, $\lambda$ the absorbed power, and the fluid density, kinematic viscosity, thermal diffusivity and thermal conductivity, respectively. \citet{Karcher2000Turbulent} experimentally investigated the flow of liquid iron ($Pr\approx0.1$) in a vacuum, heated by a high-power electron beam up to \SI{50}{\kilo\watt}. At Marangoni numbers between $10^{7}$ and $10^{8}$ they find turbulent convection, indicated by a $Nu \sim Ma^{1/3}$ scaling behavior of the dimensionless heat transfer\citep{Pumir1996Heat}. \citet{Dikshit2009Convection} experimentally observed a scaling between the laminar and turbulent behavior at Marangoni numbers between \num{e4} and \num{e5}.
\citet{Boeck2003LowPrandtlNumber} studied a similar electron beam heating process for a $Pr=0.1$ fluid via three dimensional direct numerical simulations. They find a transition from stationary to oscillatory convection around $Ma\approx2\times10^{4}$ and eventually chaotic flow above $Ma\approx2\times10^5$, and good agreement with scaling laws derived by \citet{Pumir1996Heat}. \citet{Kuhlmann2010Flow} used linear stability analysis to study the occurrence of flow instabilities in thermocapillary liquid pools in cylindrical domains with multiple aspect ratios, heated by a non-uniform heat source. For low Prandtl number fluids above a critical Marangoni number, they find a stationary centrifugal instability driven by inertial effects.

We conduct our study of the flow in a pool of a low Prandtl number liquid with surfactants ($Pr=\mathcal{O}(\num{e-1})$) with a non-uniform heat-flux at the free surface using dynamic large eddy simulations (LES). The well-resolved large eddy simulations are validated using a high resolution direct numerical simulation (DNS). We will provide a novel insight into the flow inside the liquid pool and investigate the occurrence of flow instabilities and turbulence over a range of Marangoni numbers.

\section{Mathematical Model}

\subsection{Governing equations}

\begin{figure}
\includegraphics{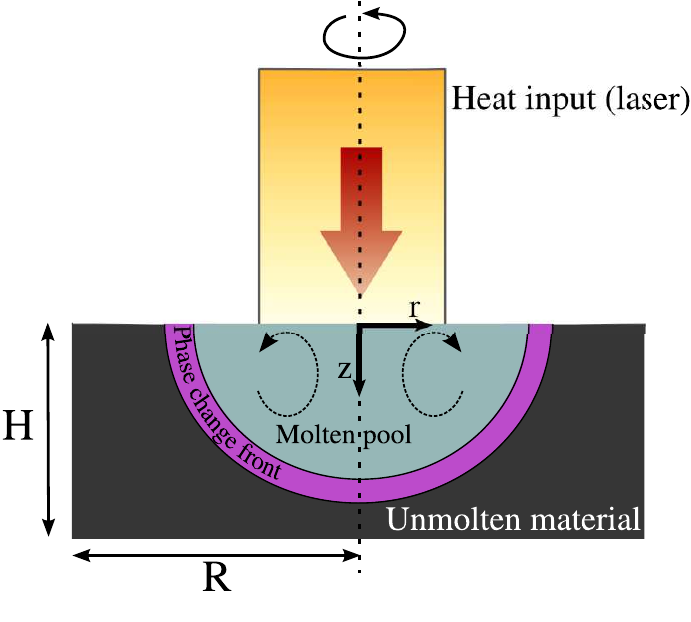}

\protect\caption{Schematic representation of the studied problem. \label{fig:pool-schema}}
\end{figure}

We investigate a solid cylinder of metal targeted on one end by a high power laser. The laser irradiation is absorbed by the metal, leading to an increase in temperature and eventually a melting phase change and the creation of a liquid melt pool. Heat is transferred into the solid bulk of the material by conduction and thermocapillary driven convection (Fig.~\ref{fig:pool-schema}). This system is mathematically modeled with an energy transport equation  with a source term for the latent heat of the phase change

\begin{equation}
\frac{D}{Dt}(T)=\nabla\cdot(\alpha\nabla T)+S_{latent}\label{eq:Energy-equation}
\end{equation}
\nomenclature[aD]{$\frac{D}{Dt}$}{Material derivative}\nomenclature[atime]{$t$}{Time}\nomenclature[grho]{$\rho$}{Density}\nomenclature[acp]{$c_p$}{Heat capacity}\nomenclature[aT]{$T$}{Temperature}\nomenclature[glambda]{$\lambda$}{Thermal conductivity}\nomenclature[aS_latent]{$S_{latent}$}{Latent heat source term}
with the thermal diffusivity $\alpha=\lambda/\rho c_{p}$.

Due to the non-uniform heating of the top surface, large radial surface temperature
gradients develop. These result in radial gradients in surface tension, leading to thermocapillary forces along the liquid-gas interface driving flow in the molten pool. The momentum transport is described by the Navier-Stokes equations, with a momentum sink to account for solid and semi-solid regions

\begin{equation}
\frac{D}{Dt}\vec{U}=\nabla\cdot(\nu\nabla\vec{U}) -\frac{1}{\rho}\nabla p -\vec{F}_{damp}\label{eq:Navier-Stokes}
\end{equation}
Here, we have assumed constant density over all phases, and have neglected buoyancy since the Grashof number $Gr = \mathcal{O}(\num{e3})$ is small \cite{Kuhlmann2010Flow}. Furthermore, since surface tension is very strong in liquid metals ($Ca = \mu{}U/\gamma = \mathcal{O}(\num{e-1})$, with $U \sim \Delta\gamma/\mu$), we assume the liquid-gas interface to be undeformable.

\nomenclature[aU]{$\vec{u}$}{Fluid velocity}\nomenclature[ap]{$p$}{Pressure}\nomenclature[gmu]{$\mu$}{Dynamic viscosity}\nomenclature[aFdamp]{$\vec{F}_{damp}$}{Momentum sink term due to solidification}


The effect of melting and solidification on the heat transfer is
taken into account via the source term $S_{latent}$ in equation \ref{eq:Energy-equation},

\begin{equation}
S_{latent}={\displaystyle \frac{h_{f}}{c_p}\frac{dg}{dt}}\label{eq:latent-heat-term}
\end{equation}
\nomenclature[aH]{$h_f$}{Latent heat of fusion}\nomenclature[ag]{$g$}{Volume fraction of solid}

with $h_f$ the specific latent heat of fusion, and $g$ the volume fraction of solid material, which is assumed to vary linearly over the melting temperature range between solidus $T_s$ and liquidus $T_l$ 

\begin{equation}
g=\frac{T_{l}-T}{T_{l}-T_{s}},\, T_{s}<T<T_{l}\label{eq:latent-heat-solid-fraction-fs}
\end{equation}

\nomenclature[aTsl]{$T_s$, $T_l$}{Solidus and liquidus temperature}


Through the inclusion of the momentum sink term, the momentum equation \ref{eq:Navier-Stokes} is valid for the entire domain including liquid, semi-solid and solid regions. The (semi-)solid regions are modeled as a porous medium, introducing a momentum sink following the isotropic Blake-Kozeny model~\citep{Singh2001Modelling}

\begin{equation}
\vec{F}_{damp}=\frac{\mu K}{\rho}\vec{U}=\frac{\vec{U}}{\rho}\mu K_{0}\frac{g^{2}}{(1-g)^{3}+\varepsilon}
\end{equation}

with $\varepsilon=10^{-3}$ and $K_0$ a constant coefficient characterizing the porosity.

\subsection{Boundary conditions}


At the top surface, the laser irradiation is modeled as a top-hat distributed heat flux. Because the heat loss from the outer surfaces of the cylindrical domain to the ambient due to radiation and convection is only a small fraction of the laser irradiation, we apply adiabatic boundary conditions everywhere except the irradiated area, where we apply a top-hat distribution as

\begin{equation}
\lambda\nabla_{n}T\Big|_{z=0}=\frac{P}{\pi r_{q}^{2}},\, r\leq r_{q}\label{eq:bc_laser_heat}
\end{equation}

\nomenclature[geta]{$\eta$}{Laser absorptivity}\nomenclature[aP]{$P$}{Laser power}\nomenclature[arq]{$r_q$}{Laser beam radius}

Here we set the laser beam radius $r_q=\SI{1.4}{\milli\meter}$ and absorbed powers P of 170, 195, 250, 500 and \SI{675}{\watt}, to mimic realistic absorbed irradiation relevant for metals processing applications \cite{Pitscheneder1996Role}.


\nomenclature[ggamma]{$\gamma$}{Surface tension}\nomenclature[xsubscriptt]{$t$}{Tangential direction}
\nomenclature[xsubscriptn]{$n$}{Normal direction}
At all surfaces except the liquid-gas interface, we set the velocity to zero. At the gas-liquid interface, which is assumed to remain flat, we impose a shear stress in the liquid due to surface tension gradients along the interface, known as Marangoni or thermocapillary stress:

\begin{equation}
\mu\nabla_{n}U_{t}\Big|_{z=0}=\frac{d\gamma}{dT}\nabla_{t}T\label{eq:Marangoni-BC}
\end{equation}

The surface tension temperature coefficient $\partial\gamma / \partial T$ has a defining influence on the nature of the flow as it determines the only driving force acting here. As discussed in the introduction, in the present work we wish to study the stability of flows in liquids containing a homogeneously distributed surface active element. The surfactant will react with the liquid metal, significantly reducing the surface tension at the liquid-gas interface. 
At increasing temperatures, chemical bonds between metal atoms and surfactant molecules break, and surfactant molecules dissociate. Hence the surface tension increases at increasing temperatures and $\partial\gamma / \partial T$ is positive. At very high temperatures however, after all surfactant molecules have dissociated, the liquid will start to behave as the pure liquid metal and $\partial\gamma / \partial T$ will approach the constant, negative value of the pure liquid metal. The transition between the states is smooth, and there is a critical temperature $T_c$ where $\partial\gamma / \partial T$ changes its sign and the surface tension has a local maximum. The surface tension gradients drive a flow from low to high surface tension areas. In laser heating applications, the maximum temperature will typically be at the center of the beam, decreasing towards its edge. A chemically pure liquid metal would thus have its highest surface tension at the edge of the pool and flow towards that cooler edge. In melts with impurities the surface tension maximum will shift towards the center of the irradiated area, and thus the resulting flow pattern will be different.

\begin{figure}
\includegraphics{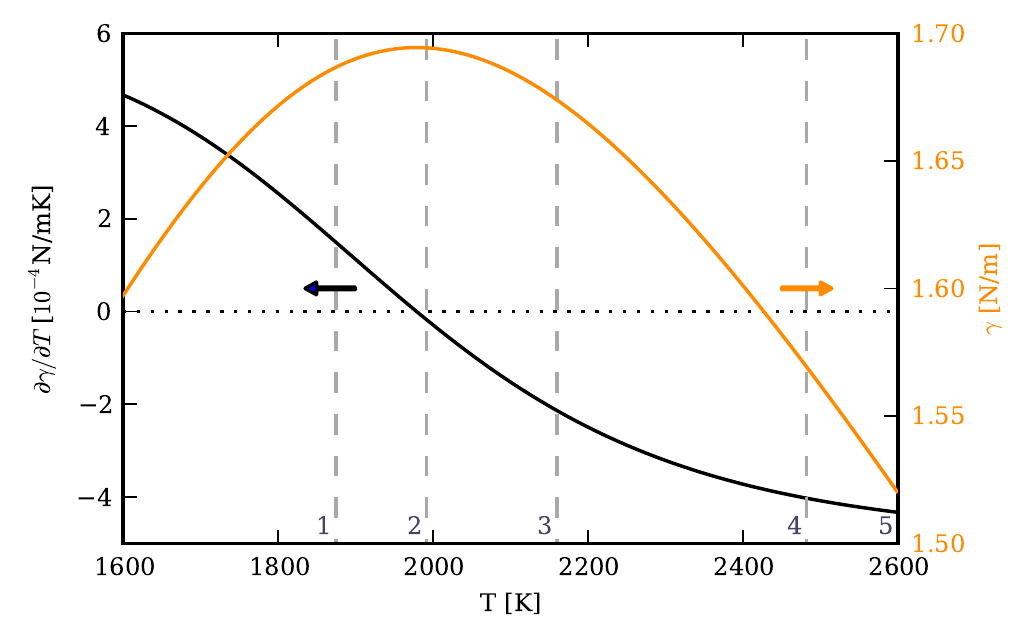}

\protect\caption{Surface tension temperature coefficient. The critical temperature at which the sign change occurs is $T_c = \SI{1979}{\kelvin}$. The vertical dashed lines indicate the maximum surface temperature reached in cases 1-5, as discussed later.\label{fig:dsdT-150ppm}}
\end{figure}

A realistic relationship \citep{Sahoo1988Surface,Hibiya2013Effect} between the surface tension temperature coefficient $\partial \gamma / \partial T$ and temperature for a binary liquid metal -- surfactant mixture, as shown in figure~\ref{fig:dsdT-150ppm}, will be used in the simulations:
\begin{equation}
\gamma=\gamma_0 - \partial\gamma/\partial T|_0 (T-T_0) - RT\Gamma_s \ln \left[1 + k_l a_s \exp(\Delta H^0 / (RT))\right]
\end{equation}
This relationship, based on the combination of Gibbs and Langmuir adsorption isotherms, has been derived by \citet{Sahoo1988Surface} for a binary iron-sulfur system. The shape of the surface tension--temperature curve and thus the system behavior is not specific to the iron-sulfur system, but also applies to e.g. silver \citep{Hibiya2013Effect} and nickel \citep{Ozawa2014Influence} in the presence of oxygen.


\subsection{Non-dimensional formulation}

Using the non-dimensional variables $x^* = x/L$, $t^* = t\alpha/{L^2}$, $\vec{U}^* = \vec{U} L/\alpha$, $T^* = T/\Delta T$, $p^*=p L^2/\rho\alpha^2$, following \citet{Pumir1996Heat}, and $K^*=L^2K$ the governing equations can be written as
\begin{equation}
\frac{1}{Pr}\left(\frac{\partial \vec{U}^*}{\partial t^*}+\nabla\cdot(\vec{U}^*\vec{U}^*)+\nabla p^*\right)=\nabla^2\vec{U}^* + K^* \vec{U}^*
\end{equation}

\begin{equation}
\frac{\partial T^*}{\partial t^*} + \nabla \cdot (\vec{U}^*T^*) = \nabla^2 T^* + \frac{1}{St}\frac{\partial g}{\partial t^*}
\end{equation}

and the boundary conditions

\begin{equation}
\nabla_n U^*_t = Ma \nabla_t T^*
\end{equation}

The dimensionless numbers appearing in the governing equations are the Prandtl number $Pr=\nu/\alpha=0.18$ and the Stefan number $St=c_p\Delta T/h_f$, gauging the ratio of sensible to latent heat. In the following, we will take the laser beam radius $r_q$ as the characteristic length scale $L$, and $P/(L\lambda)$ as the characteristic temperature difference $\Delta T$. With this, and for constant $\partial \gamma / \partial T$, the Marangoni number $Ma={\partial\gamma}/{\partial T}{L\Delta T}({\mu\alpha})^{-1}$, i.e. the ratio of surface tension to viscous forces, is commonly defined as
\begin{equation}
Ma=\frac{\partial\gamma}{\partial T} \frac{P}{\mu\alpha\lambda}
\end{equation}
For the present case with varying $\partial\gamma/\partial T$, we linearize $\partial\gamma/\partial T$ around $T_c$ leading to $\partial\gamma/\partial T \sim (\partial^2\gamma/\partial T^2)|_{T_c}\Delta T$ and define $Ma$ as
\begin{equation}
Ma = \frac{\partial^2\gamma}{\partial T^2}\bigg|_{T_c} \frac{\Delta T^2 L}{\mu\alpha} \sim \frac{\partial^2\gamma}{\partial T^2}\bigg|_{T_c} \frac{P^2}{\mu\alpha\lambda^2 L}
\end{equation}
with $\partial^2\gamma / \partial T^2|_{T_c}=\SI{1.38e-6}{\newton\per\meter\per\kelvin\squared}$. The resulting dimensionless numbers are tabulated in Tab.~\ref{tab:dimless-numbers}. The relevant material properties are listed in Tab.~\ref{tab:Material-properties} in the appendix. The ratio $Ma/Gr$ is very large indeed, which further justifies neglecting buoyancy forces.

\begin{table}
\protect\caption{Dimensionless numbers\label{tab:dimless-numbers}}
\begin{ruledtabular}
\begin{tabular}{l c c c}
Case & P [\si{\watt}] & Ma & St \tabularnewline
\midrule
1 & 170 & \num{2.1e6}{} &  4.7  \tabularnewline
2 & 195 & \num{2.8e6}{} &  5.3  \tabularnewline
3 & 250 & \num{4.6e6}{} &  6.7  \tabularnewline
4 & 500 & \num{1.8e7}{} & 13.5  \tabularnewline
5 & 675 & \num{3.4e7}{} & 18.3  \tabularnewline

\end{tabular}
\end{ruledtabular}
\end{table}

\subsection{Discretization}

Our solver has been developed using the open source finite volume framework OpenFOAM (version 2.1.x) \citep{Weller1998Tensorial}. The time derivative is discretized with a second order backward differencing scheme. The divergence terms in the transport equations are discretized with the second order limitedLinear(V) TVD scheme \citep{Berberovic2010Investigation}. At every time step, the non-linearity associated with the pressure-velocity-coupling is handled by the iterative PISO algorithm \citep{Issa1986Solution}. Once a divergence free velocity field has been computed at a given time step, the temperature equation is solved. The non-linearity of the temperature equation due to latent heat is dealt with using an implicit source term linearization technique \citep{Voller1991GENERAL}. 

\begin{figure}
\includegraphics{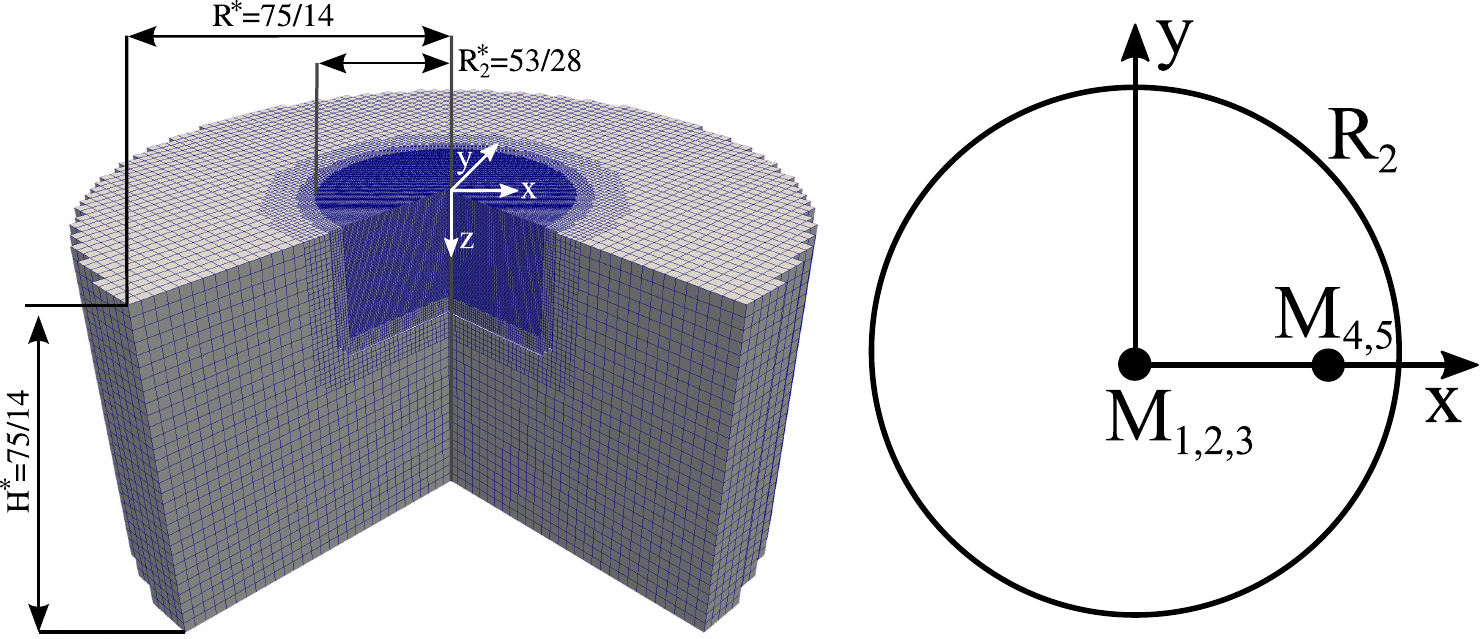}

\protect\caption{3D mesh, where one quarter of the domain has been clipped for visualization. The coarse outer mesh with a grid cell size of 
 $\Delta{}x^*=0.178$ is refined in three steps to the finest inner mesh with a grid cell size of 
 $\Delta{}x^*=0.022$. The latter is too fine to be resolved in this figure. The sketch on the right shows monitoring point locations at a depth of $z^*=3/14$ and radial distances of $x^*=0$ ($M_{1,2,3}$) and $x^*=15/14$ ($M_{4,5}$).\label{fig:3D-mesh}}
\end{figure}

The solution domain is a cylinder of radius $R=\SI{7.5}{\milli\meter}$ and height $H=\SI{7.5}{\milli\meter}$ (see Fig.~\ref{fig:pool-schema}), discretized with a mesh of 2.15 million cubic control volumes. The region where we expect fluid flow consists of small cubes with a a cell spacing of \SI{31}{\micro\meter}, whereas we use larger cells of \SI{250}{\micro\meter} away from the liquid region. The mesh is shown in figure~\ref{fig:3D-mesh}. The time step is dynamically set obeying a maximum Courant number of $Co=U\Delta t/\Delta x<0.35$, resulting in a typical time step of roughly \SI{15}{\micro\second}. A simulation on 48 cores (Intel E5-2650 v2) of the LISA compute cluster\footnote{\protect\url{https://surfsara.nl/systems/lisa}} is completed in roughly four weeks time.

To model turbulence, we use Large Eddy Simulations based on the dynamic Smagorinsky approach as proposed by Lilly \citep{Lilly1992Proposed,PassalacquaDynamicSmagorinsky}.
We do not allow for backscatter of energy from the small to large scale, and thus the subgridscale viscosity $\nu_{SGS}$ is clipped to zero. The turbulent thermal diffusivity is not determined via the dynamic Smagorinsky approach, but via Reynolds analogy with $Pr_{t}=0.4$ \citep{Eidson1985Numerical,Kenjeres2006LES}.

\section{Results and discussion}

The high power laser irradiation melts the target material and within fractions of a second a liquid pool develops. The developing shape of this pool is tightly connected to the thermocapillary flow due to surface tension gradients at the free surface. These surface tension gradients in turn are dependent on the thermal gradients and the local temperature at the free surface itself. 

Preceding the discussion of the physics of our results, we present a comparison of temperature data obtained with DNS and LES for validation, as well as the subgridscale viscosity of the LES simulation, to prove the adequate performance of our simulations.

We then present the flow patterns observed on the free surface, similarly to what an experimentalist might see observing the opaque pool from the top using e.g.  PIV\citep{Zhao2009Complex,Zhao2010Effect} or PTV methods applied to particles floating on the liquid surface, followed by a discussion of the observations in the next section. In addition to the static images provided here, movies are being made available online via the journal's electronic supplement.

\subsection{Validation of LES}

\begin{figure}
\includegraphics[width=0.7\textwidth]{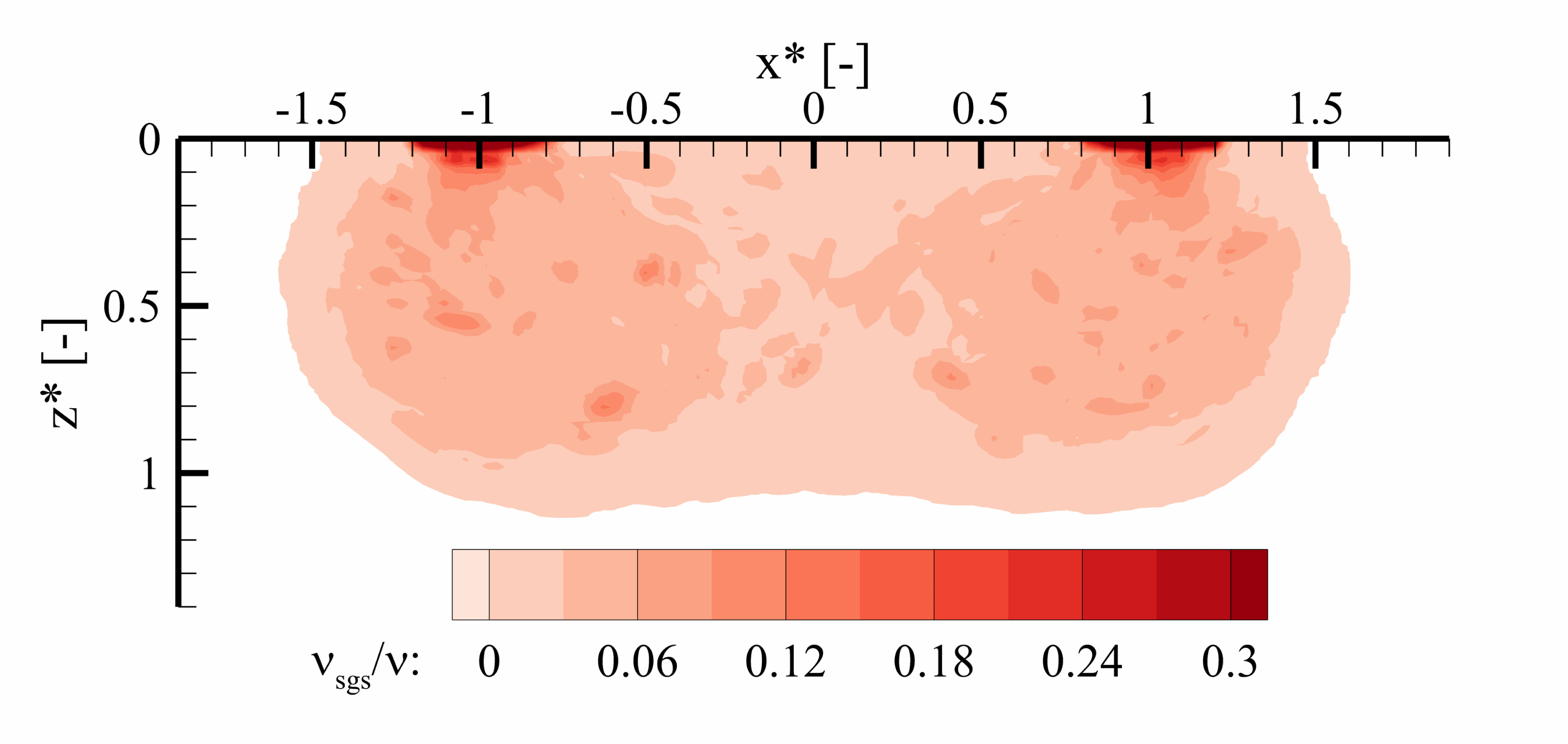}\protect\caption{Maximum values of the ratio of subgridscale and molecular diffusivity $\nu_{SGS}/\nu$ in case 4, proving a well-resolved LES simulation.\label{fig:Ratio-nusgs-nu}}
\end{figure}

Figure~\ref{fig:Ratio-nusgs-nu} shows the maximum values of $\nu_{SGS}/\nu$ obtained during the course of a simulation of case 4. The maximum value of the subgridscale turbulent viscosity $\nu_{SGS}$ is at most in the order of the molecular viscosity, which indicates a well resolved LES solution. The results obtained for the other cases are not shown here, but are similar or better.

\begin{figure}
\includegraphics[width=0.98\textwidth]{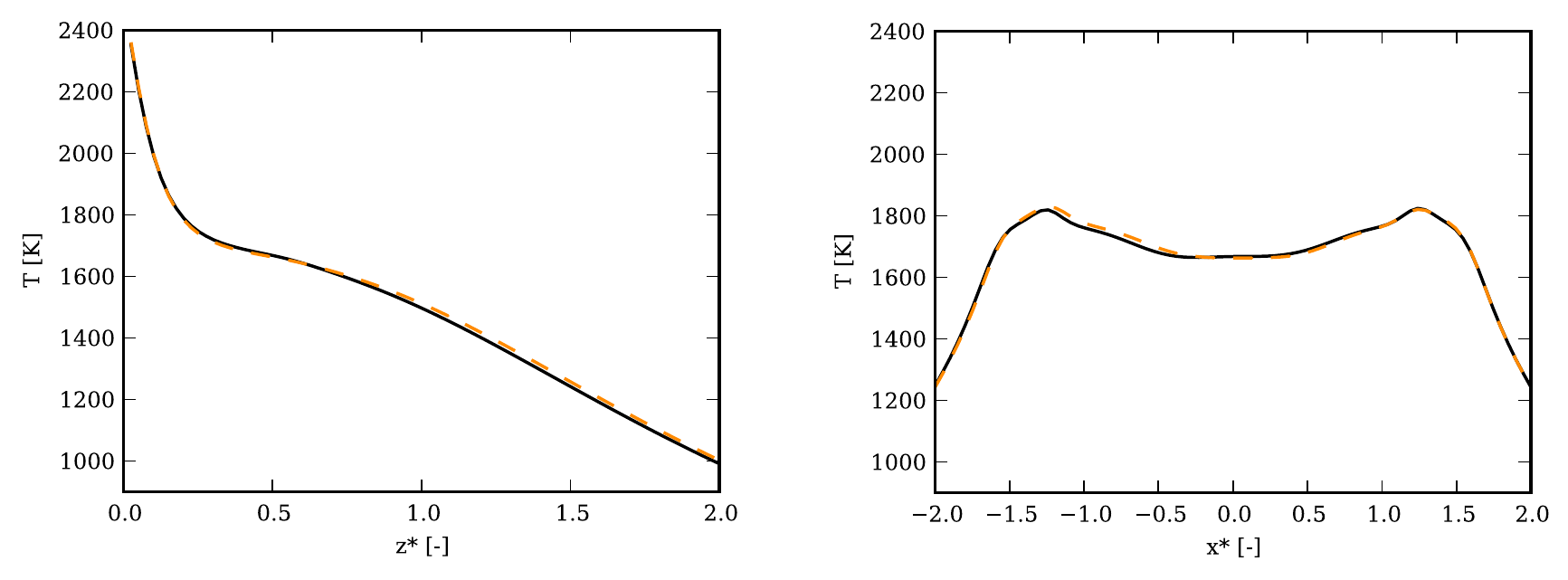}

\protect\caption{Comparison of temperatures on $x^*=0$ and $z^*=0.5$ averaged over the period from 1.5 to \SI{3.5}{\second}, computed by dynamic LES (dashed, 2.5M cell mesh) and DNS (solid, 21M cell mesh).\label{fig:dles-dnsf}}
\end{figure}

To further validate our results, we have performed a direct numerical simulation for the highest laser power (case 5) on a grid with 21 million cells (\SI{14}{\micro\meter} grid spacing). Time averaged temperature profiles over a period of \SI{2}{\second} are shown in figure~\ref{fig:dles-dnsf}.
We can conclude adequate performance of the dynamic LES solution on a 10x coarser mesh. It is reasonable to expect the solution to be similarly good or better for the other laser powers.

\subsection{Free surface flow}

\begin{figure}
\subfloat[Case 1\label{fig:Unsteady-flow-field-13-FS}]{\includegraphics[height=0.325\textwidth,angle=270,trim=0cm 0cm 0cm 1.7cm, clip]{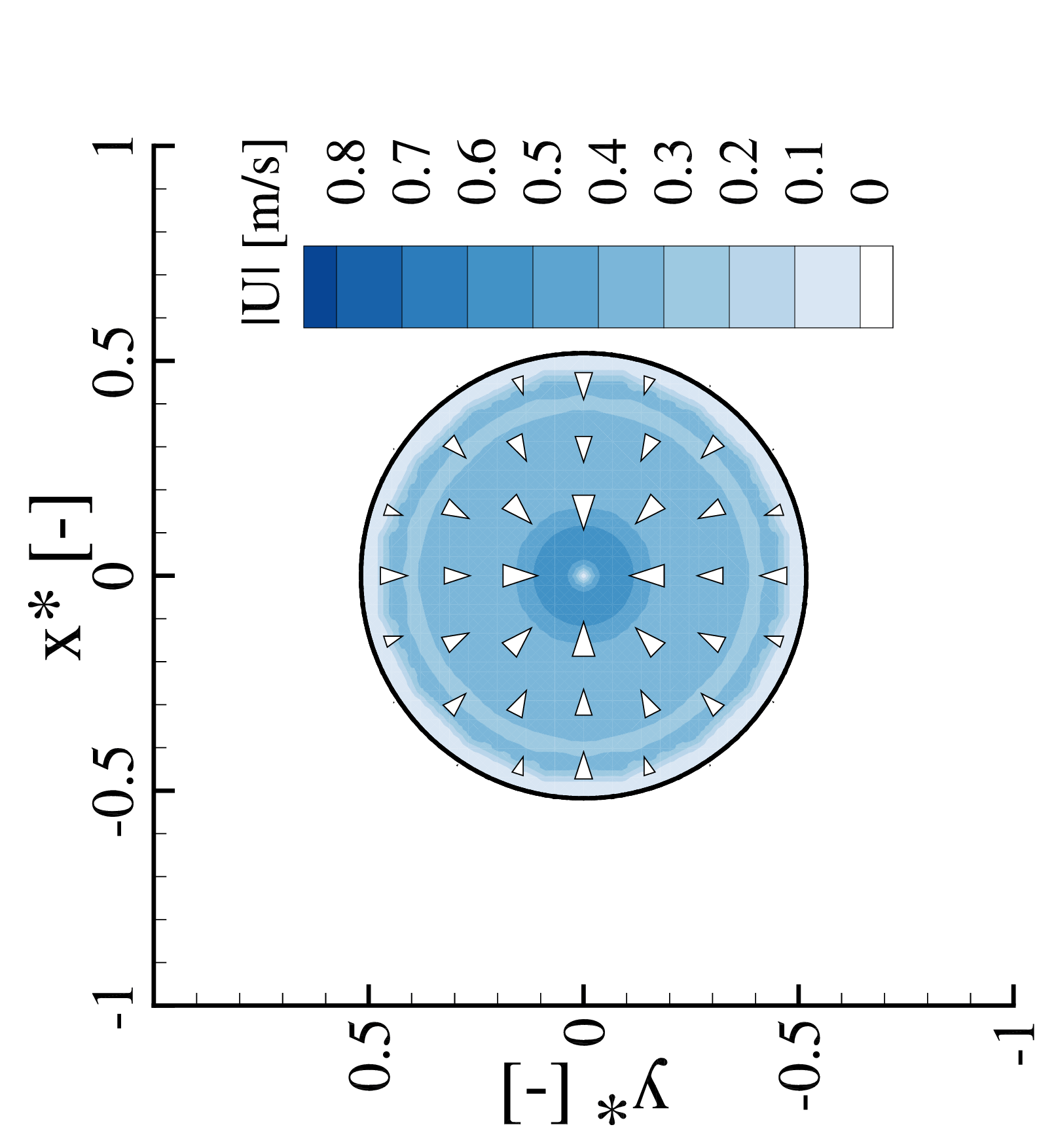}
} 
\subfloat[Case 2\label{fig:Unsteady-flow-field-15-FS}]{\includegraphics[height=0.3\textwidth,angle=270,trim=0cm 1cm 0cm 1.7cm, clip]{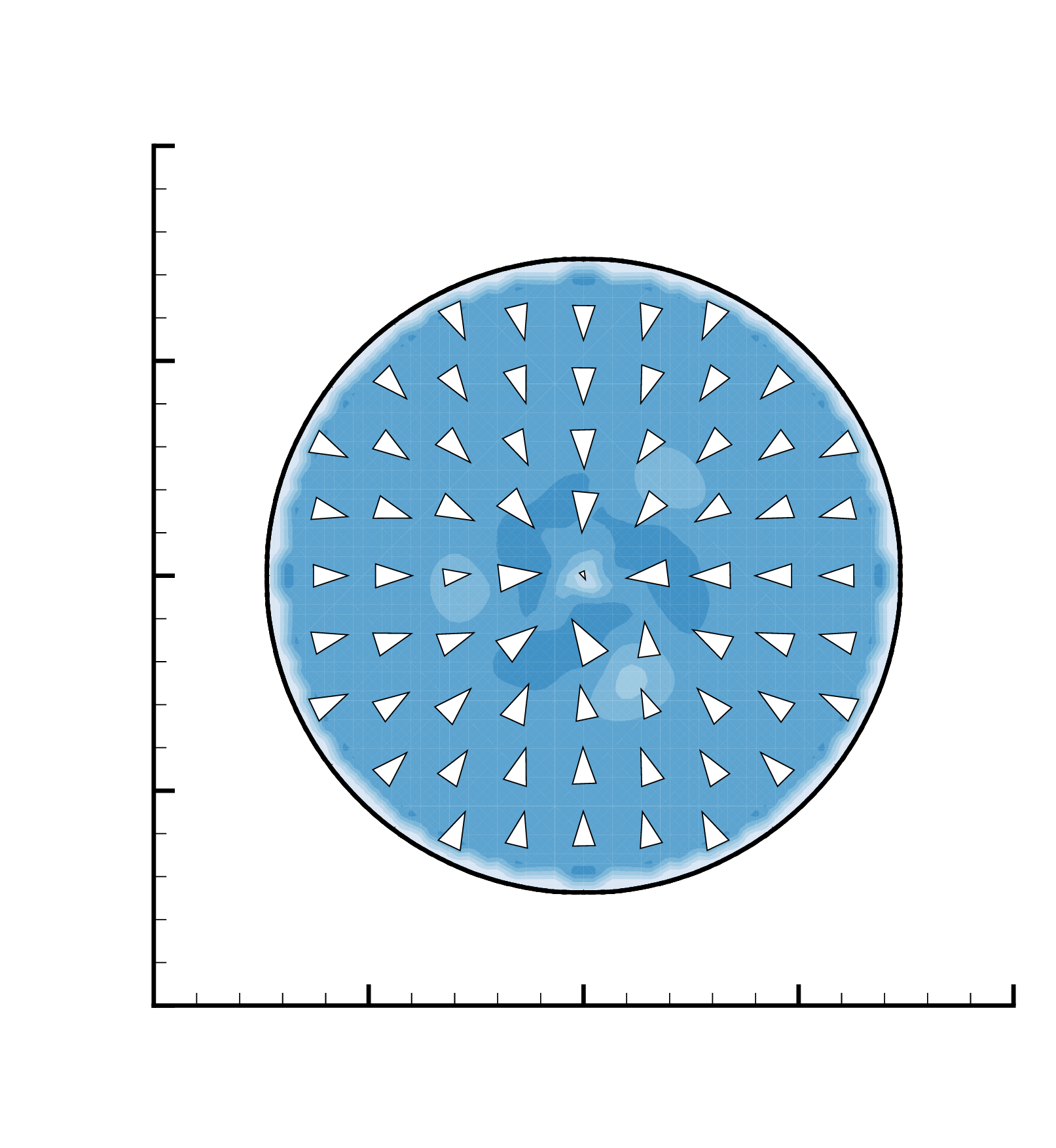}
} 
\subfloat[Case 3\label{fig:Unsteady-flow-field-19-FS}]{\includegraphics[height=0.3\textwidth,angle=270,trim=0cm 1cm 0cm 1.7cm, clip]{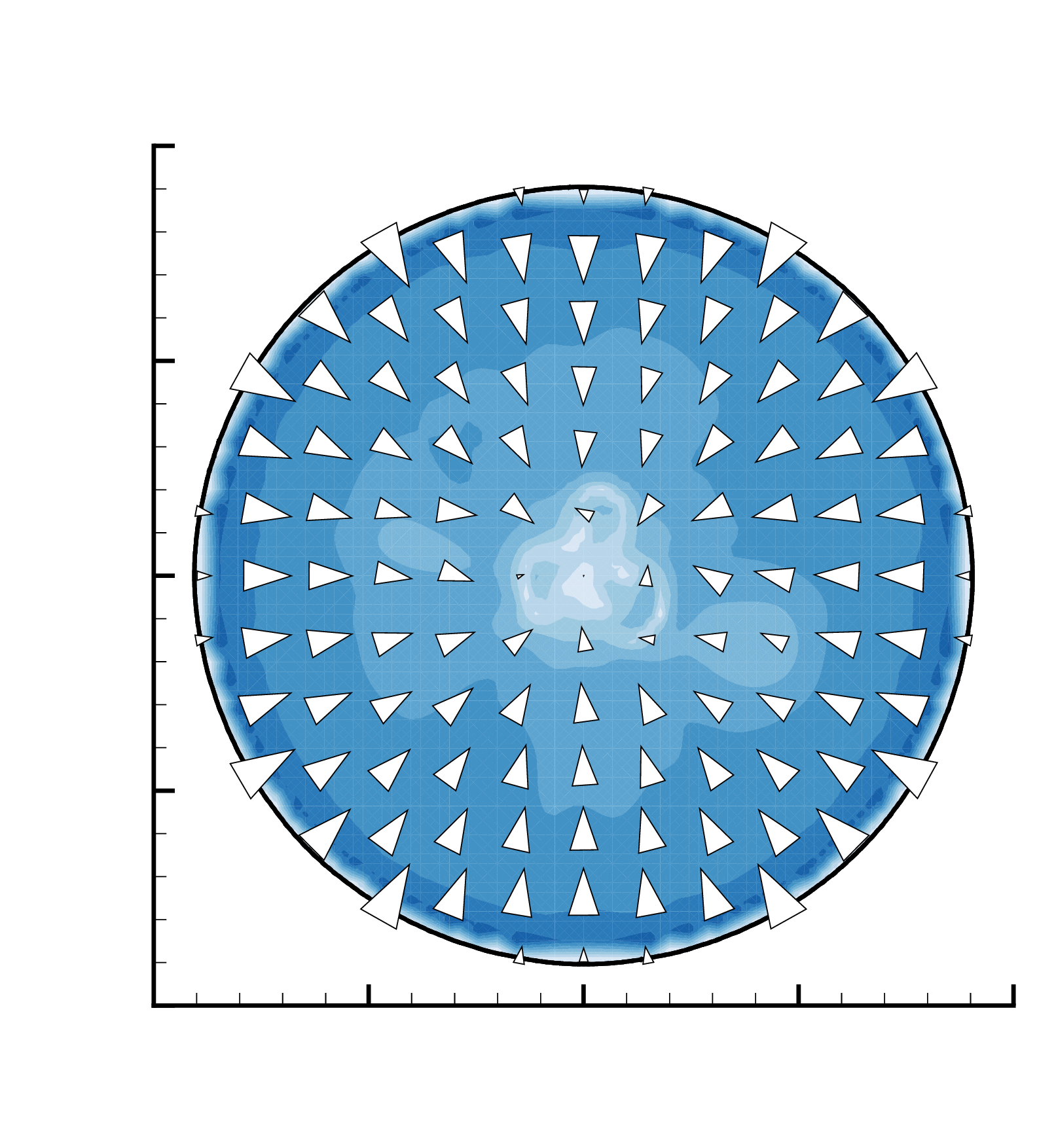}
} \\
\subfloat[Case 4\label{fig:Unsteady-flow-field-38-FS}]{\includegraphics[height=0.45\textwidth,angle=270,trim=0cm 0cm 0cm 0cm, clip]{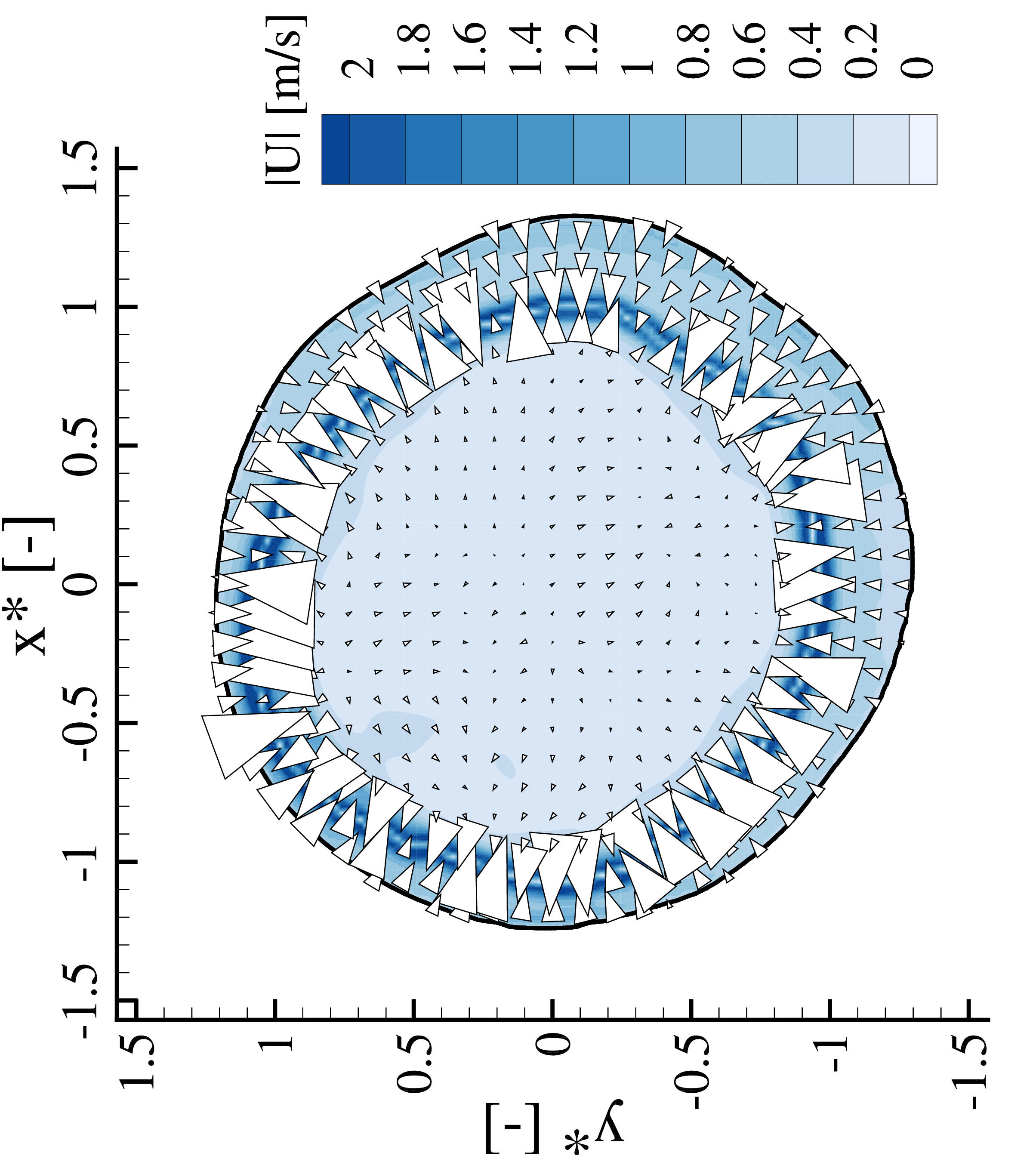}
}
\subfloat[Case 5\label{fig:Unsteady-flow-field-52-FS}]{\includegraphics[height=0.45\textwidth,angle=270,trim=0cm 1cm 0cm 0cm, clip]{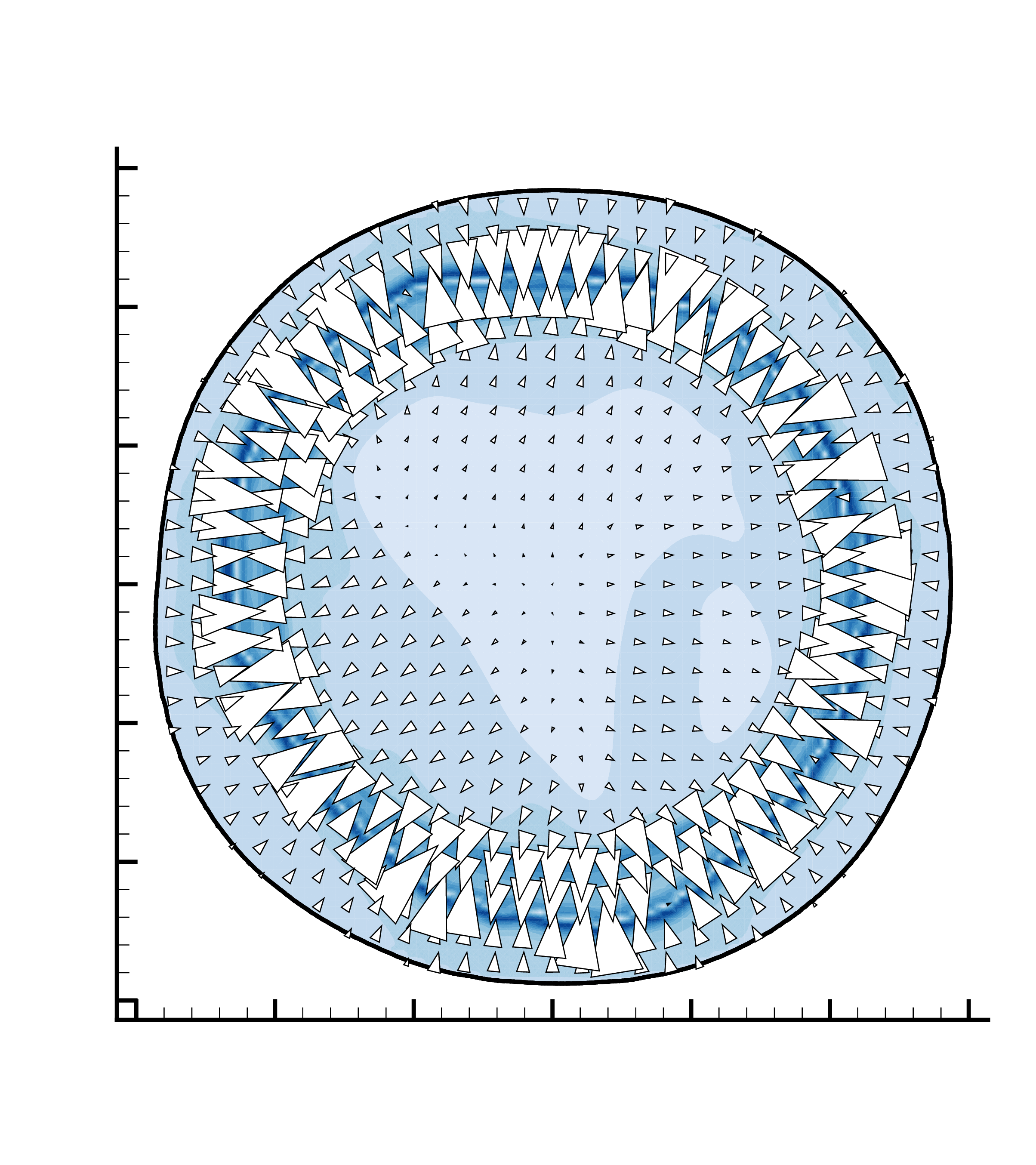}
}

\protect\caption{Free surface flow with velocity vectors and contours of velocity magnitude, at t=2s for increasing laser power: (a) Case 1, (b) Case 2, (c) Case 3, (d) Case 4, and (e) Case 5 (movies online)
}
\end{figure}

At the lowest laser power (case 1, Tab.~\ref{tab:dimless-numbers}), the flow is radially symmetric, directed from the rim of the pool towards the center, with the lowest velocities close to the pool rim gradually increasing towards the center of the pool (Fig.~\ref{fig:Unsteady-flow-field-13-FS}). The flow is stable, and even when perturbed numerically quickly restores symmetry. With increased laser power (case 2), initially the same flow pattern is observed. At some point in time, the symmetry breaks (Fig.~\ref{fig:Unsteady-flow-field-15-FS}), and a stable rotational motion develops near the center of the pool surface. Later on, a high frequency pulsating motion (radial oscillation) is superimposed onto the rotation, which remains very stable nonetheless. The same basic flow topology is observed for case 3, where the flow is again directed from the rim of the pool towards the center, though now the highest velocities are observed close to the pool rim gradually decreasing towards the center of the pool (Fig.~\ref{fig:Unsteady-flow-field-19-FS}). Again, the flow is initially axisymmetric, but very soon becomes asymmetric and unstable. The velocity distribution on the free surface shows a rotational motion. Unlike the results at the lower powers, the rotational motion is not stable and unpredictably changes its direction. For all three cases the pool boundary remains almost perfectly circular, despite any instabilities in the flow. 

The numerical experiment at a higher laser power (case 4) exhibits a dramatically different flow pattern (Fig.~\ref{fig:Unsteady-flow-field-38-FS}). Here, there is flow from the pool rim towards the pool center, which, after only a short distance from the rim, meets a flow stemming from the pool center towards the rim in a vicinity where the highest velocity magnitudes are encountered. The flow is highly unstable and even the pool boundary is significantly distorted. At the highest laser power (case 5), the flow pattern shown in Fig.~\ref{fig:Unsteady-flow-field-52-FS} is similar to the one observed in case 4. Here also an inward flow from the pool rim meets an outward flow from the pool center, leading to a stagnation region fairly close to the pool boundary. In contrast to the previous case, the stagnation region is spread over a wider area, and the pool boundary is not quite as unstable.

\subsection{Temperature profile on the free surface}

The surface flow presented in the previous section is of course driven by surface tension gradients due to non-uniform temperatures at the free surface, and thus a look at the temperature distributions on the free surface is necessary to understand the observations.

\begin{figure}
\subfloat[Square pattern\label{fig:15-isotherm-square}]{\includegraphics[height=\textwidth,angle=270,trim=0cm 0cm 0cm 0.75cm,clip]{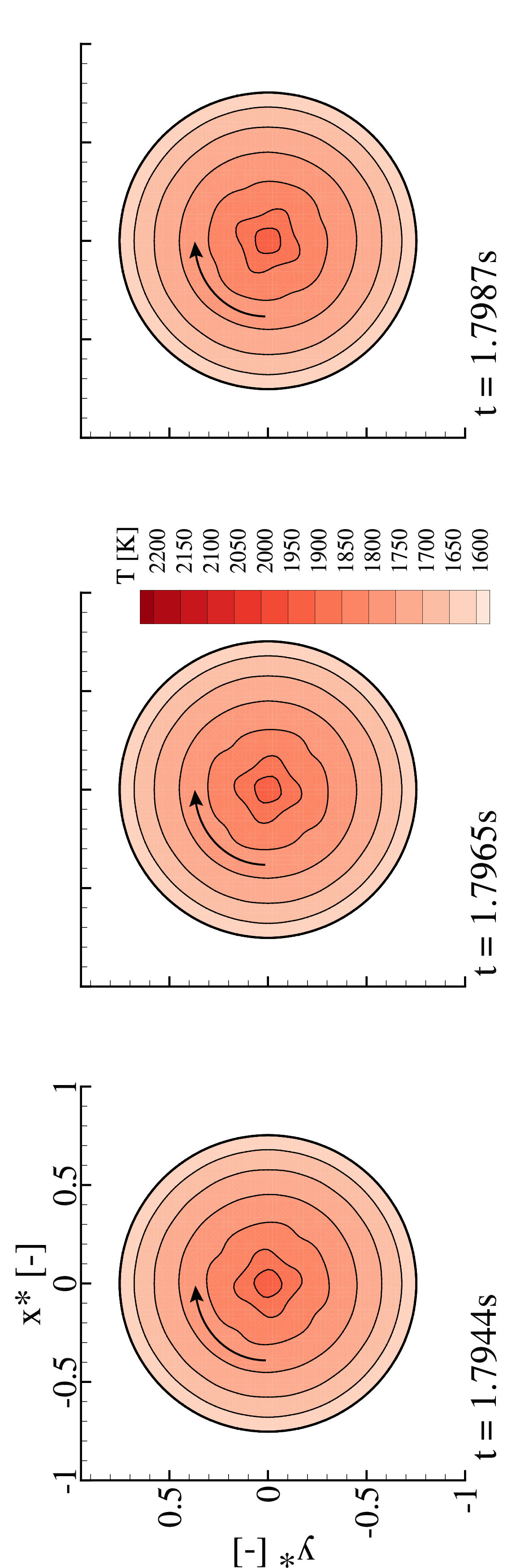}
}
\\
\subfloat[Trefoil pattern\label{fig:15-isotherm-trefoil}]{\includegraphics[height=\textwidth,angle=270,trim=0cm 0cm 0cm 0.75cm,clip]{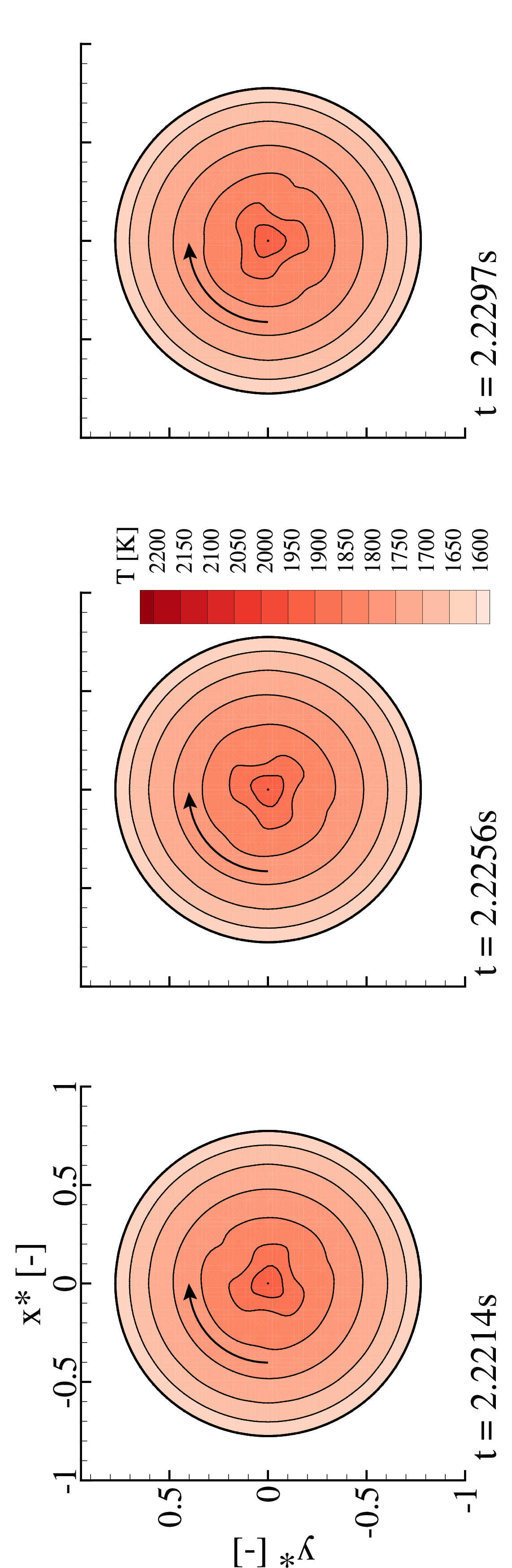}
}
\protect\caption{Steady clockwise rotational motion of initially square and later trefoil shaped isotherm pattern on free surface, case 2 (movie online).}
\end{figure}

\begin{figure}
\subfloat[Clockwise rotation\label{fig:19-isotherm-clockwise}]{\includegraphics[height=\textwidth,angle=270,trim=0cm 0cm 0cm 0.75cm,clip]{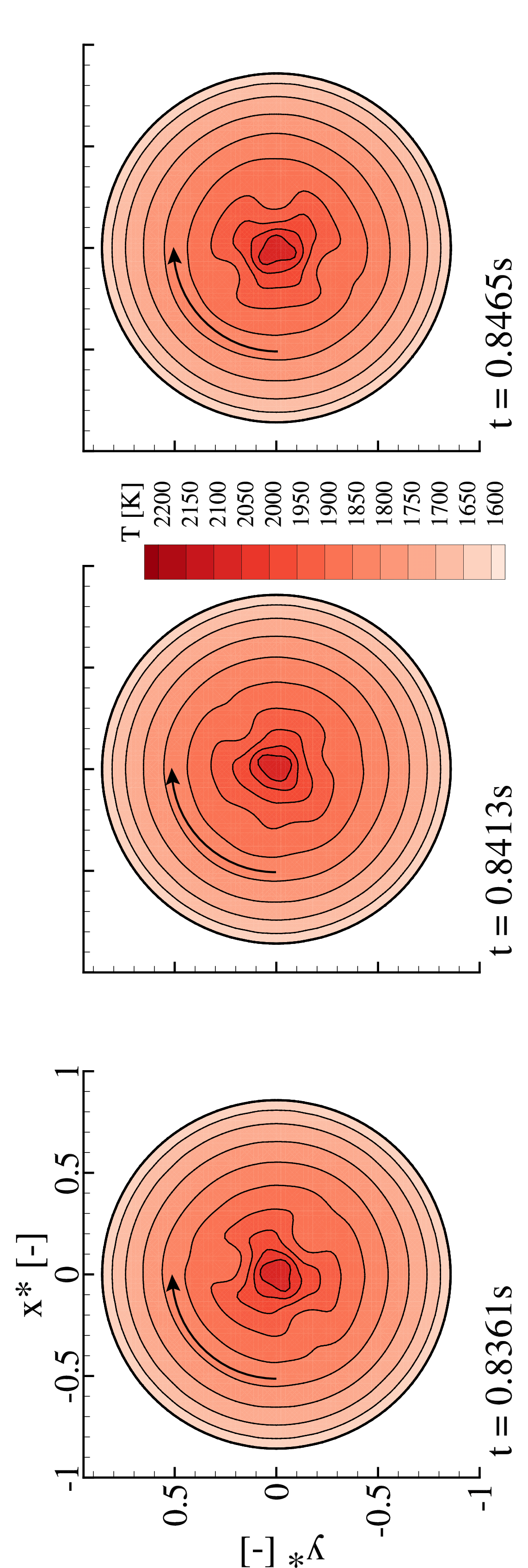}
}\\
\subfloat[Counter-clockwise rotation\label{fig:19-isotherm-counterclock}]{\includegraphics[height=\textwidth,angle=270,trim=0cm 0cm 0cm 0.75cm,clip]{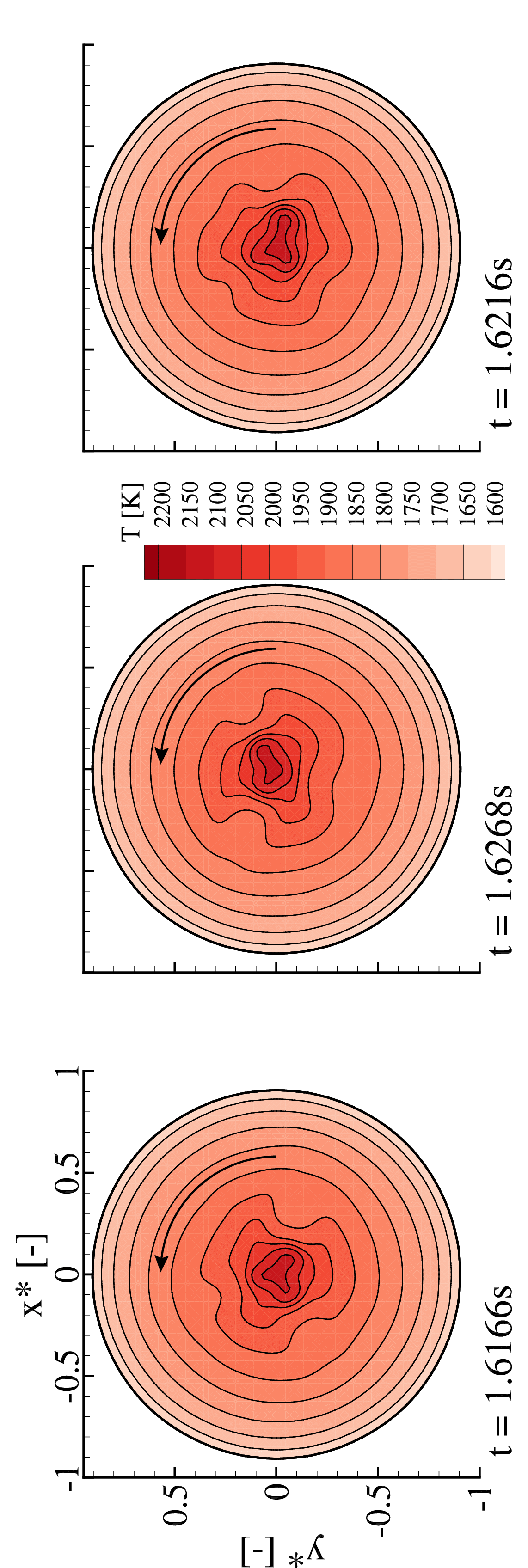}
}
\protect\caption{Isotherms on free surface, case 3, showing clockwise rotation of the trefoil pattern at a certain time instance followed by counter-clockwise rotation at a later time (movie online).}
\end{figure}

At the lowest laser power (case 1, Fig.~\ref{fig:dsdT-150ppm}), the temperature distribution is radially symmetric with a maximum temperature in the center well below $T_c=\SI{1979}{\kelvin}$. The surface force is thus always directed towards the hot pool center. Increasing the laser power (case 2), the initially similar symmetric pattern is superseded by first a rotating square (Fig.~\ref{fig:15-isotherm-square}) and then a rotating trefoil pattern (Fig.~\ref{fig:15-isotherm-trefoil}). The rotational instability develops before the maximum temperature at the surface reaches $T_c$. Once the maximum temperature surpasses $T_c$, a pulsating motion superimposes the rotation. Higher laser powers are necessary to push the temperature significantly beyond $T_c$. In case 3, temperatures up to \SI{2300}{\kelvin}, well beyond $T_c$, are sustained at the free surface in a narrow region near the pool center, while the majority of the free surface temperature remains below $T_c$. This leads to a sign change in $\partial\gamma/\partial T$ (cf. Fig.~\ref{fig:dsdT-150ppm}), which explains the flow pattern shown previously (Fig.~\ref{fig:Unsteady-flow-field-19-FS}), where the flow directed towards the center of the pool is decelerated by the opposing Marangoni force but not permanently reversed. The strongest Marangoni forces appear close to the pool boundary, due to a combination of the highest thermal gradients and the highest value of $\partial\gamma/\partial T$ in this region. Qualitatively, the surface temperature patterns are similar to case 2 with an initial radially symmetry of the isotherms, which breaks down much sooner compared to case 2. The isotherms in the vicinity the pool center then briefly deform into a square shape, followed by a trefoil shape which rotates with time (Fig.~\ref{fig:19-isotherm-clockwise}). Unlike case 2, the rotation is stable only for short periods, and unpredictably reverses its direction (Fig.~\ref{fig:19-isotherm-counterclock}). A qualitatively similar asymmetrical rotational surface flow has previously been observed in an experimental study of laser melting of steel exposed to surface active oxygen \citep{Zhao2010Effect,Zhao2009Complex}.

\begin{figure}
\subfloat[Case 4\label{fig:38fs-isotherms}]{\includegraphics[height=\textwidth,angle=270,trim=0cm 0cm 0cm 0.75cm,clip]{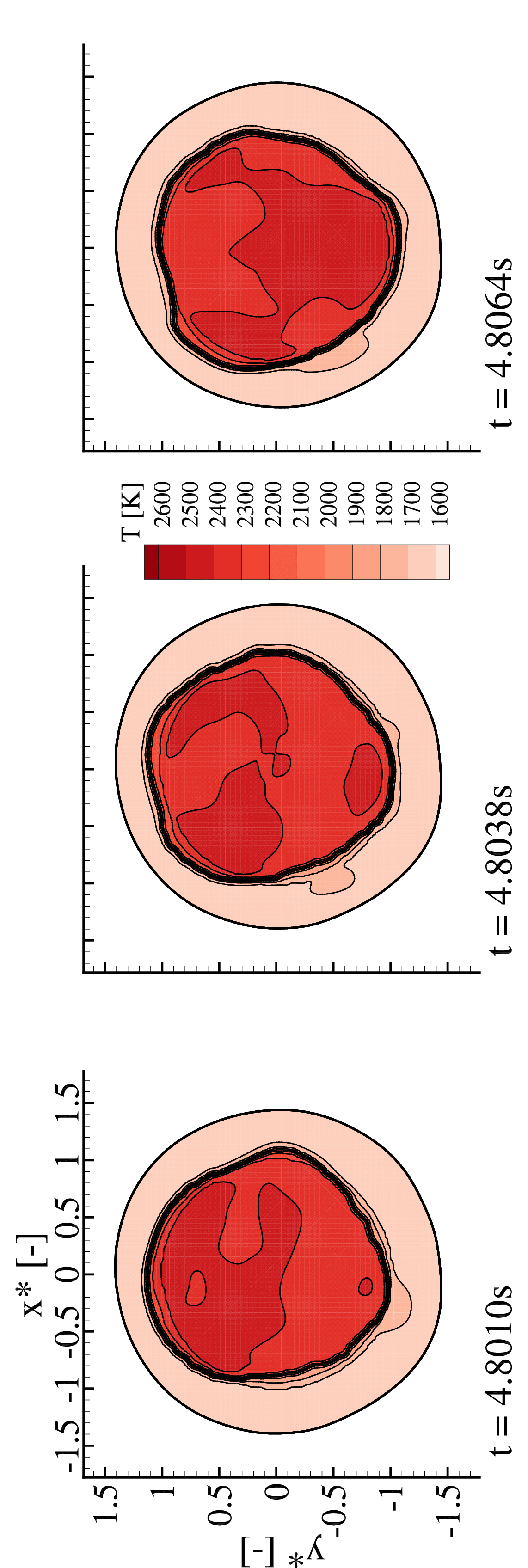}
}\\
\subfloat[Case 5\label{fig:52fs-isotherms}]{\includegraphics[height=\textwidth,angle=270,trim=0cm 0cm 0cm 0.75cm,clip]{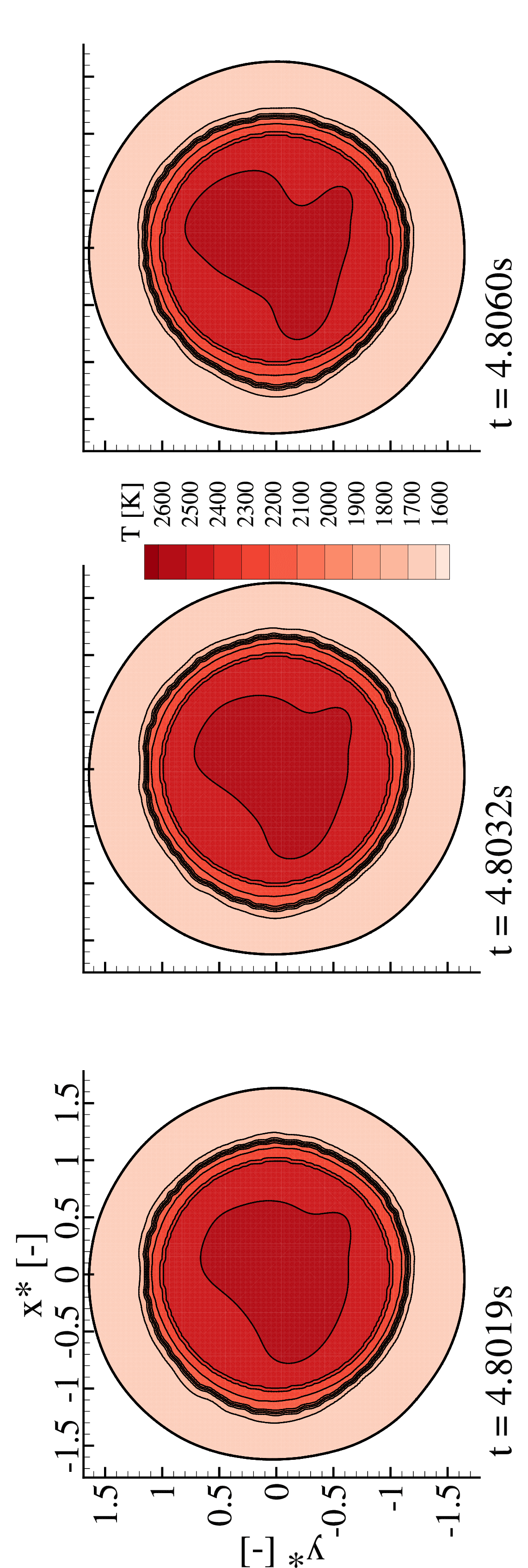}
}
\protect\caption{Isotherms on the free surface for case 4 and 5, showing highly chaotic temporal behavior for case 4 and irregular, less intermittent behavior for case 5 (movies online).}
\end{figure}

When the laser power is increased even further, the resulting higher temperatures increases the strength of the Marangoni flow directed towards the pool boundary, and the region with temperatures above $T_c$ is larger. As a result, the flow topology changes because a flow from the pool center to the boundary is now sustainable. For the second largest power, case 4, maximum temperatures of around \SI{2600}{\kelvin} are observed. The outward flow shown earlier in Fig.~\ref{fig:Unsteady-flow-field-38-FS} drives heat from the pool center towards the pool boundary. As a result, the temperatures around the center of the pool are fairly uniform, whereas extreme temperature gradients of \SI{4500}{\kelvin\per\milli\meter} are encountered in the stagnation region. Since the stagnation region is only $\approx\SI{0.1}{\milli\meter}$ wide, small spatial disturbances result in large changes in the temperature gradient and thus the resulting local Marangoni forces. This makes the flow highly unstable, and strong disturbances are indeed visible on the free surface of the pool (Fig.~\ref{fig:38fs-isotherms}), which evolve rapidly in time.

At the highest laser power, case 5 (Fig.~\ref{fig:52fs-isotherms}), the pool volume is larger, but the general flow features are similar to the previously discussed case 4. The stagnation region is quite a bit wider, leading to much lower gradients of \SI{2100}{\kelvin\per\milli\meter}, and the oscillations observed at the free surface are less intermittent.

With the computed temperatures at the free surface, we can determine Nusselt numbers $Nu = P/(r_q\lambda (T_{max}-T_s))$. The values for all cases lie between 14 and 22, highlighting the large effect of fluid flow on heat transfer in the liquid pools.

\begin{figure}
\includegraphics[width=0.65\textwidth]{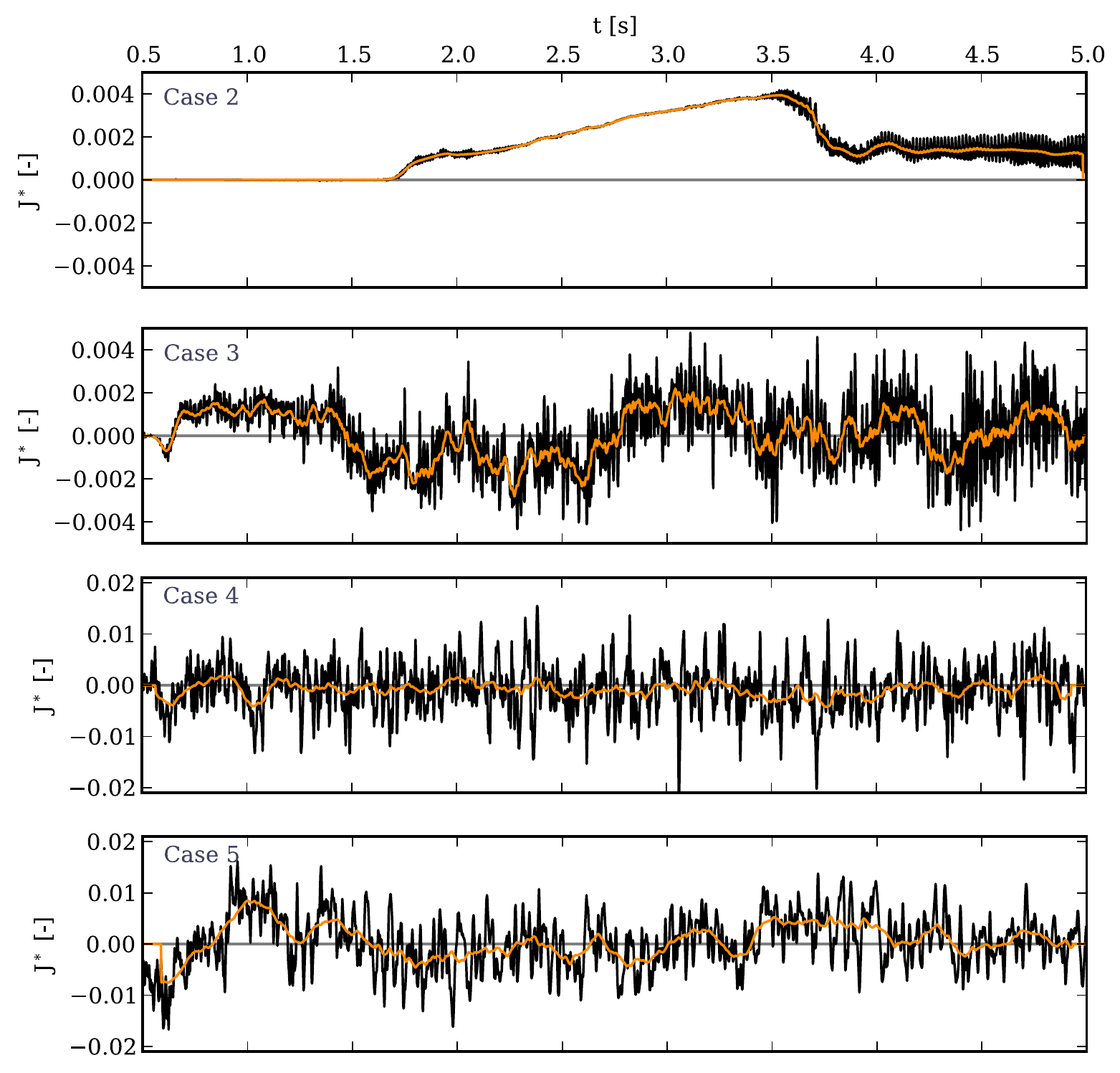}
\protect\caption{Surface averaged angular momentum $J^*$ of the free surface over time (orange line: moving average) for increasing laser power. Negative values translate to counter-clockwise motion.\label{fig:angMom}}
\end{figure}

There is a clearly visible rotational motion of the temperature pattern on the free surface in cases 2 and 3. Judging plainly from the temperature profile at the free surface such a motion could not be identified in the temperature profiles at the higher laser powers. We have therefore investigated the temporal behavior of the spatially averaged, normalized angular momentum $J^*(t)=\langle\vec{r}\times\vec{U}\rangle_A/\langle\vec{r}\cdot \vec{U}\rangle_A=\langle-yu_{x}+xu_{y}\rangle_{A}/\langle x u_x + y u_y \rangle_{A}$, where $\langle\rangle_{A}$ denotes the average over the pool surface\nomenclature[aJ]{$J$}{Angular momentum} (Fig.~\ref{fig:angMom}). For the stable non-rotational flow in case 1, $J^*(t)=0$ for all $t$ (not shown here). The onset of stable clockwise rotation after \SI{1.7}{\second} in case 2, and the onset of pulsation after roughly \SI{3.6}{\second} is clearly distinguished from the initial stable flow. Also clear are the reversals of the direction of rotation occurring for case 3. For case 4, the angular momentum over time frequently changes its sign and no clear rotational motion is visible in the plot. The averaged angular momentum at the highest laser power, case 5, exhibits more persistent features than the chaotic case 4, but the rotation is not as clear as in cases 2 and 3.

\begin{figure}
\includegraphics[width=0.65\textwidth]{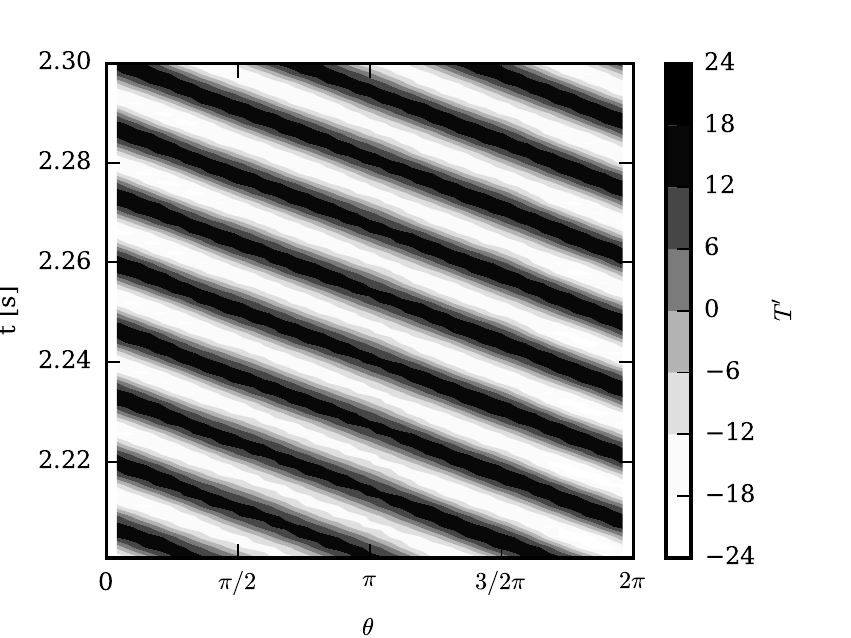}
\protect\caption{Temperature fluctuation $T'=T- \frac{1}{2\pi}\oint_0^{2\pi} T  d\theta$ at $r^*=0.143$ in case 2, with $\theta$ the azimuthal coordinate}\label{fig:tfluc15}
\end{figure}

The rotationally oscillating surface temperature patterns and the oscillating rotational momentum of the pool surface reported here, particularly at low power (case 2), bear resemblance to hydrothermal waves \citep{SmithDavis1983Instabilities,Davis1987Thermocapillary,Lappa2010Thermal}, which have previously also been observed for capillary driven flows in low Prandtl number liquids in annular configurations \citep{LiImaishi2007Bifurcation}. At low Prandtl numbers, these waves travel obliquely to the thermal gradient. For the radial thermal gradients in the studied cases, this leads to thermal waves with an azimuthal velocity component, as is indeed observed. This is further illustrated in Fig.\ref{fig:tfluc15}, showing temperature perturbations around the time-averaged temperature as a function of time as a function of the azimuthal coordinate on a curve of constant r*=0.143 in case 2, resembling patterns reported for hydrothermal waves.

We do not, however, believe that hydrothermal waves are the main cause for the flow instabilities and turbulence observed experimentally in weld pools at higher laser powers \citep{Zhao2010Effect,Karcher2000Turbulent}. In simulations at high laser powers, flow instabilities leading to turbulence are present even when the flow is restricted to a 2D axisymmetric domain \citep{Kidess2016Marangoni}, excluding the presence of obliquely traveling thermal waves. The true mechanism underlying the flow instabilities will be further discussed in the following sections IIID and IIIE.

\subsection{Flow within the pool}

\begin{figure}[!ht]
\subfloat[Case 2\label{fig:15-cpl-flow}]{\includegraphics[height=0.80\textwidth,angle=270,trim=0.4cm 0cm 0cm 0cm,clip]{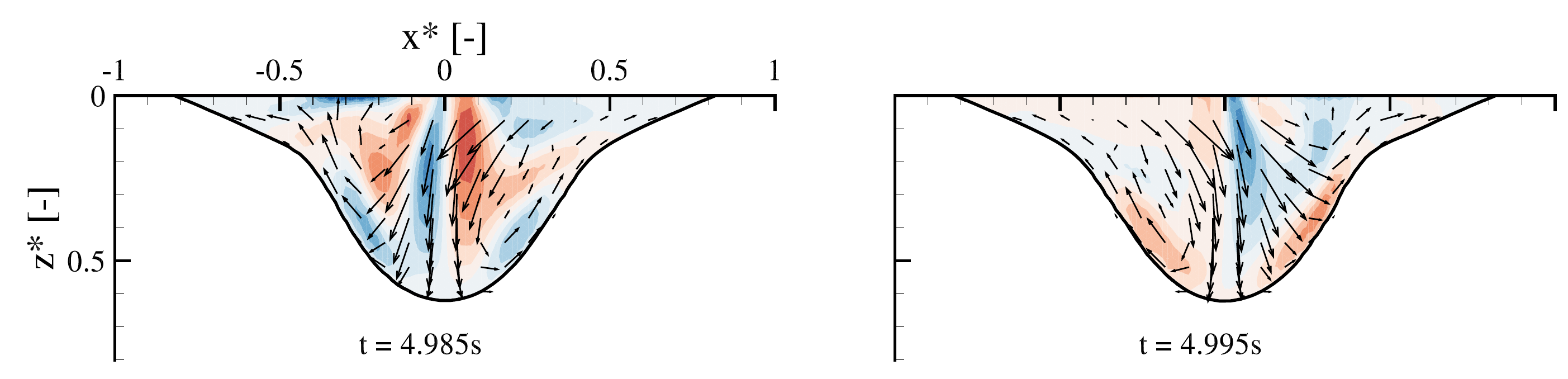}
} \\
\subfloat[Case 3\label{fig:19-cpl-flow}]{\includegraphics[height=0.80\textwidth,angle=270,trim=0.4cm 0cm 0cm 0cm,clip]{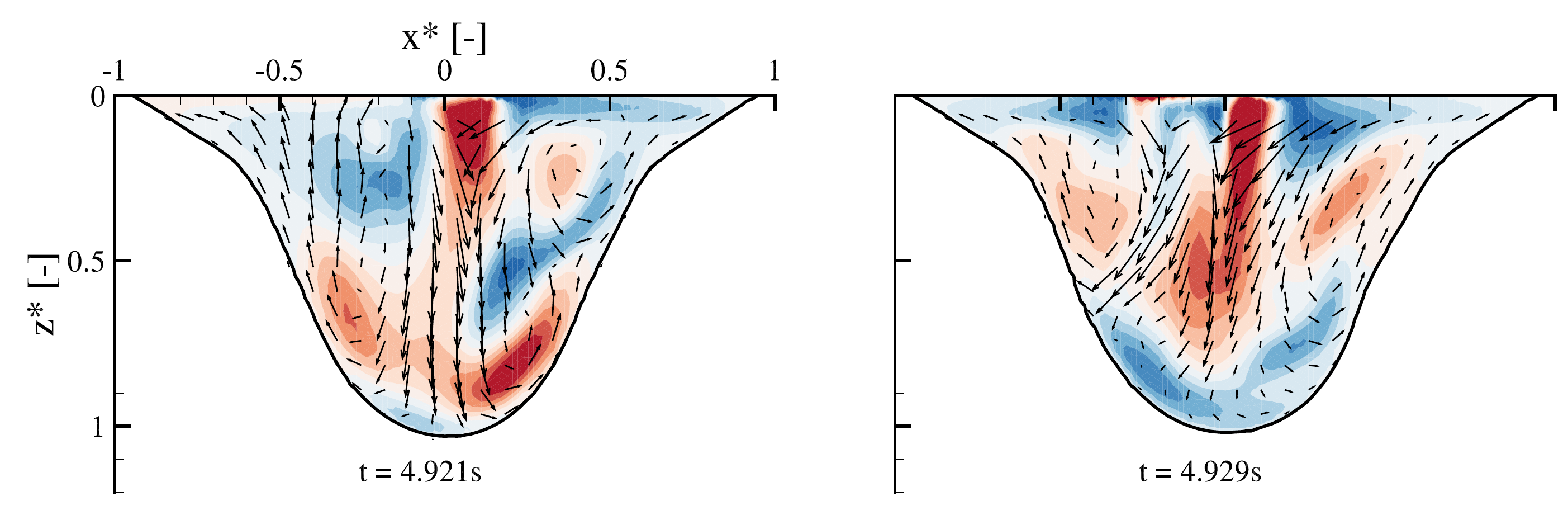}
} \\
\subfloat[Case 4\label{fig:38-cpl-flow}]{\includegraphics[height=0.92\textwidth,angle=270,trim=0.4cm 0cm 0cm 0cm,clip]{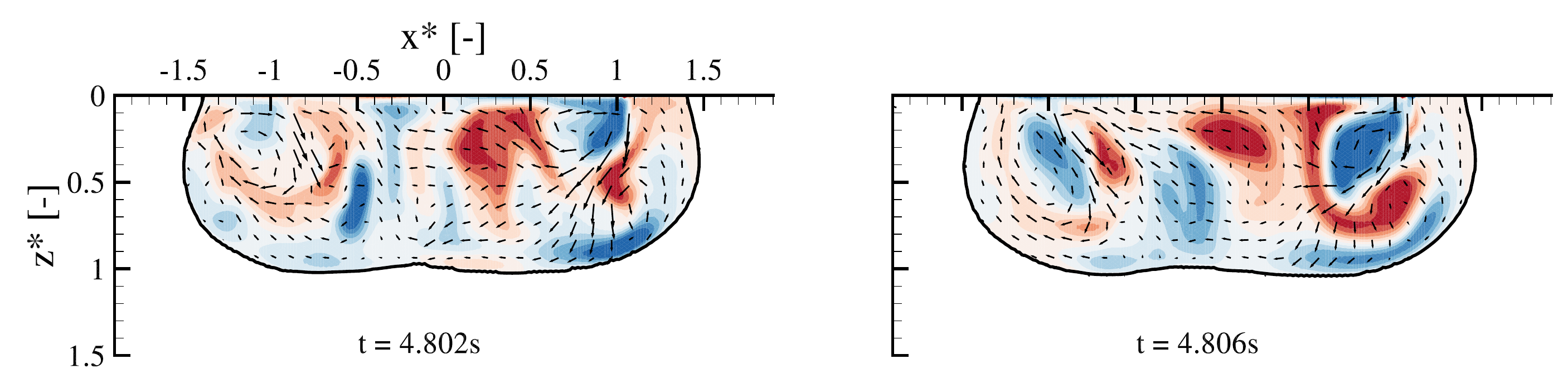}
} \\
\subfloat[Case 5\label{fig:52-cpl-flow}]{

\shortstack{
   \includegraphics[height=0.92\textwidth,angle=270,trim=0.4cm 0cm 0cm 0cm,clip]{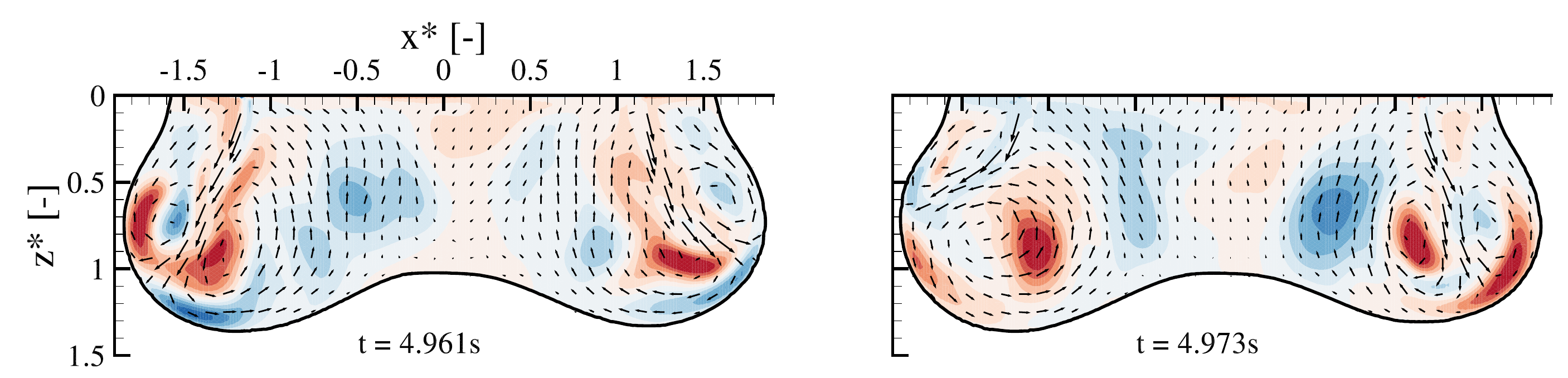} \\
   \includegraphics[height=0.6\textwidth,angle=270]{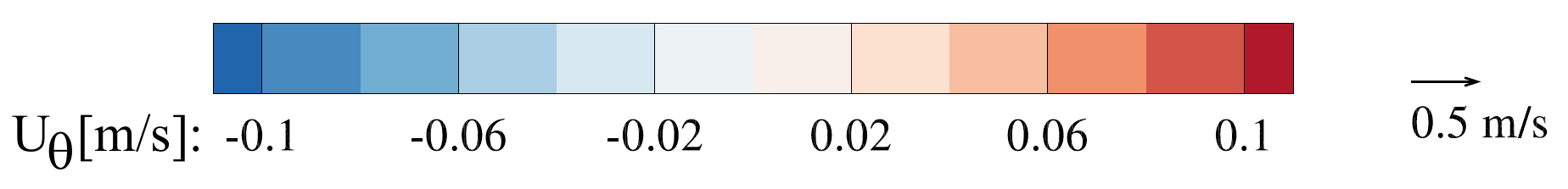}
   }
}

\protect\caption{Instationary in-plane (x,z) velocities in the $y^*=0$ plane at two time instances indicated by vectors, and azimuthal velocities indicated by color contours: (a) Case 2, (b) Case 3, (c) Case 4, and (d) Case 5 (movies online).}
\end{figure}

For the lowest laser power (case 1, not shown), the surface tension force towards the center of the pool drives a stable, symmetric donut-shaped vortex with upward flow at the edge and downward flow at the center of the pool. This leads to increased melting under the stagnation point at the center of the free surface. No azimuthal velocities are present.  
At increased laser powers the flow does not remain axisymmetric, with significant azimuthal velocities (case 2 and 3, Fig.~\ref{fig:15-cpl-flow} and Fig.~\ref{fig:19-cpl-flow}). In the cross-section of the vortex-ring, the two halves oscillate and compete, one growing at the expense of the other. The oscillation is regular in case 2, and unpredictable in case 3, where even larger, non-homogeneous azimuthal velocities are present. Even though the flow is highly unsteady, the pool boundary still remains quasi-steady.

\begin{figure}
\subfloat[Case 4\label{fig:38osci}]{\includegraphics[height=0.5\textwidth,angle=270]{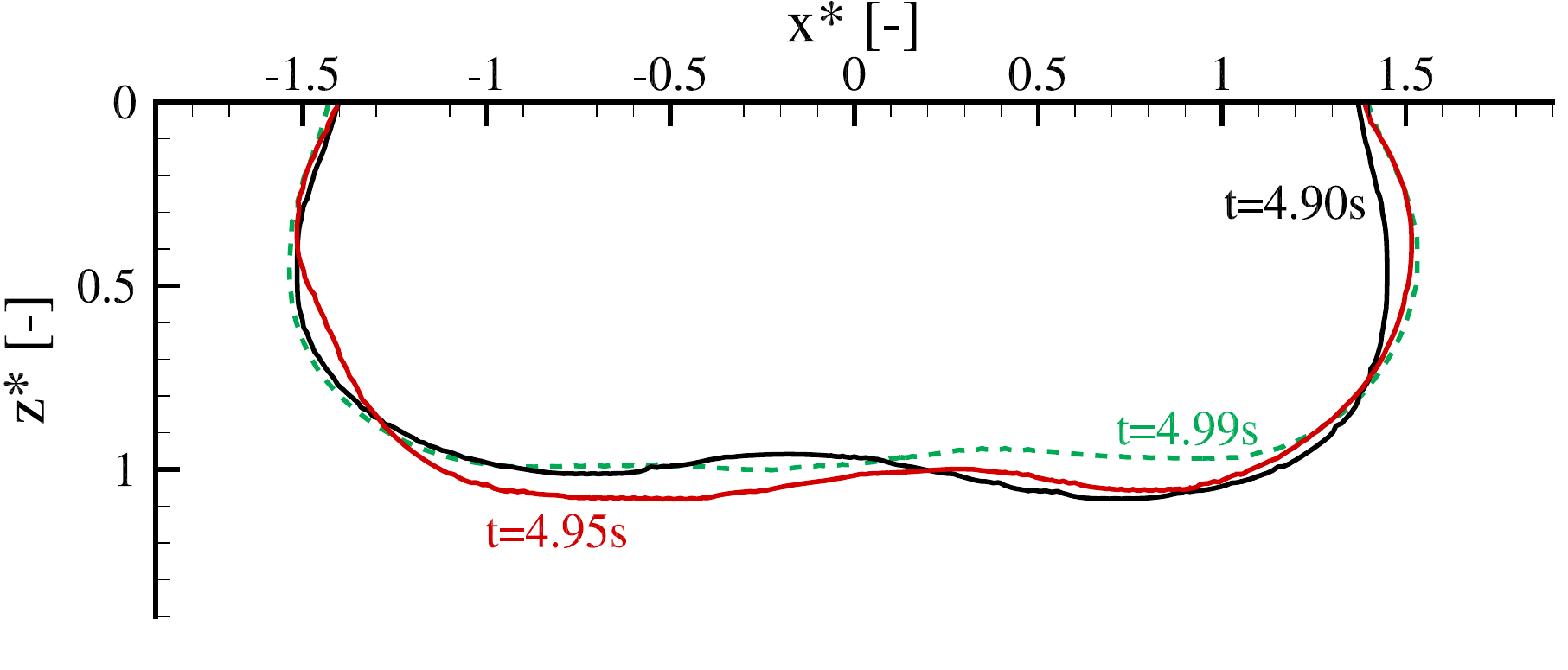}
}
\subfloat[Case 5\label{fig:52osci}]{\includegraphics[height=0.5\textwidth,angle=270]{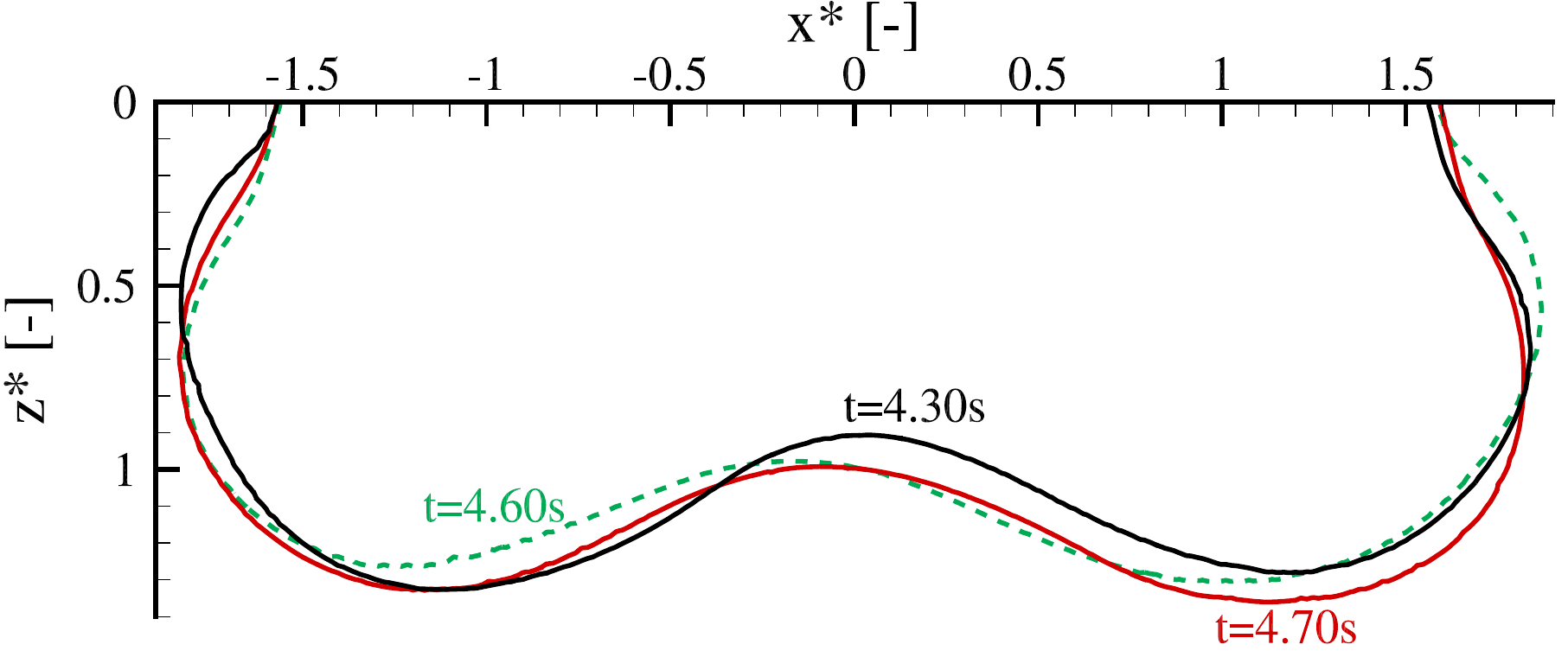}
}

\protect\caption{Pool boundary oscillation observed in case 4 and 5. The oscillation frequency is significantly higher in case 4.}
\end{figure}

For the highest laser powers, instead of the singular surface stagnation point at the lower powers, there is a stagnation ring at the free surface, where the outward surface flow from the pool center meets the inward flow from the pool boundary, from which a strong circular jet towards the base of the pool emerges. There are thus two main counter-rotating donut-shaped vortex rings which are visible as four major counter-rotating vortices in the cross-section of the pool. At early times, the outer vortex ring is small, yet exerts a big influence on the stability of the flow. A plot of the instantaneous flow obtained for case 4,  in and normal to the cross-section plane, is shown in Fig.~\ref{fig:38-cpl-flow}. We see that the flow is strongly asymmetric, and changes quickly and unpredictably within \SI{0.004}{\second}. Unlike the previously shown results at lower laser powers, not only the flow, but also the pool boundary is unsteady (Fig.~\ref{fig:38osci}). The chaotic flow in the pool leads to a uniform transfer of heat from the surface to the base of the melt pool, and thus a much more uniform pool depth. At the highest laser power of case 5 (Fig.~\ref{fig:52-cpl-flow}), the circular jet stemming from the stagnation ring between the counter-rotating vortices is stronger and somewhat less chaotic, leading to a W-shaped melt pool. The pool boundary responds to the jet oscillation similarly to case 4, however, here the boundary oscillation occurs at a lower frequency (Fig.~\ref{fig:52osci}).

To visualize the fluid flow within the opaque melt pool, an experimentalist would have to resort to X-ray radiography techniques using non-melting (e.g. tungsten) tracer particles. The particle traces will be complex two-dimensional projections of the real particle trajectory \citep{Arata1987,Czerner2005Schmelzbaddynamik,Gatzen2011Xray}, and reconstruction of the 3D particle motion is not possible unless the pool is imaged with multiple sources and detectors simultaneously \citep{Morisada2011Threedimensional}. 

\begin{figure}
\includegraphics[angle=270,width=0.45\textwidth]{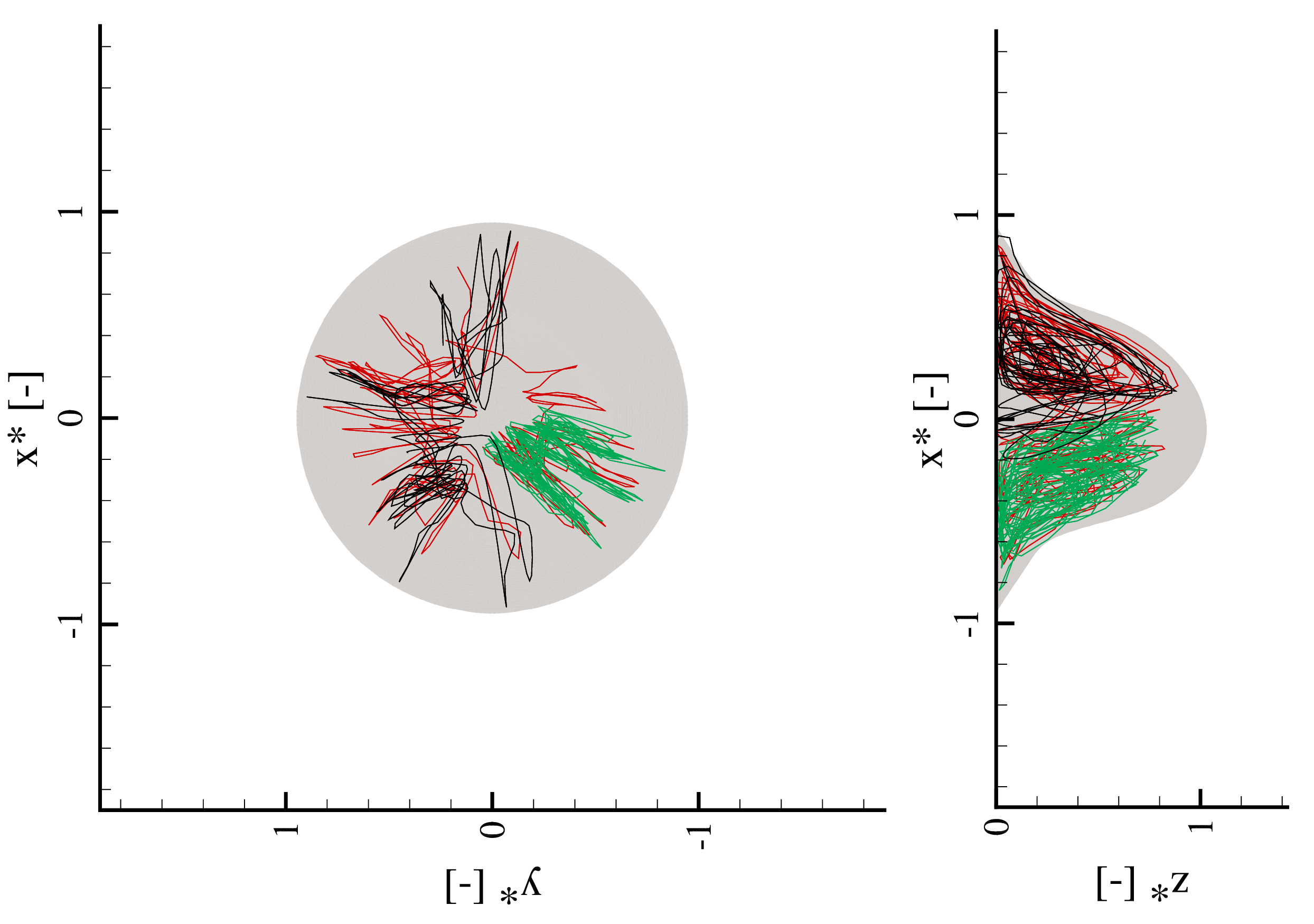}
\includegraphics[angle=270,width=0.45\textwidth]{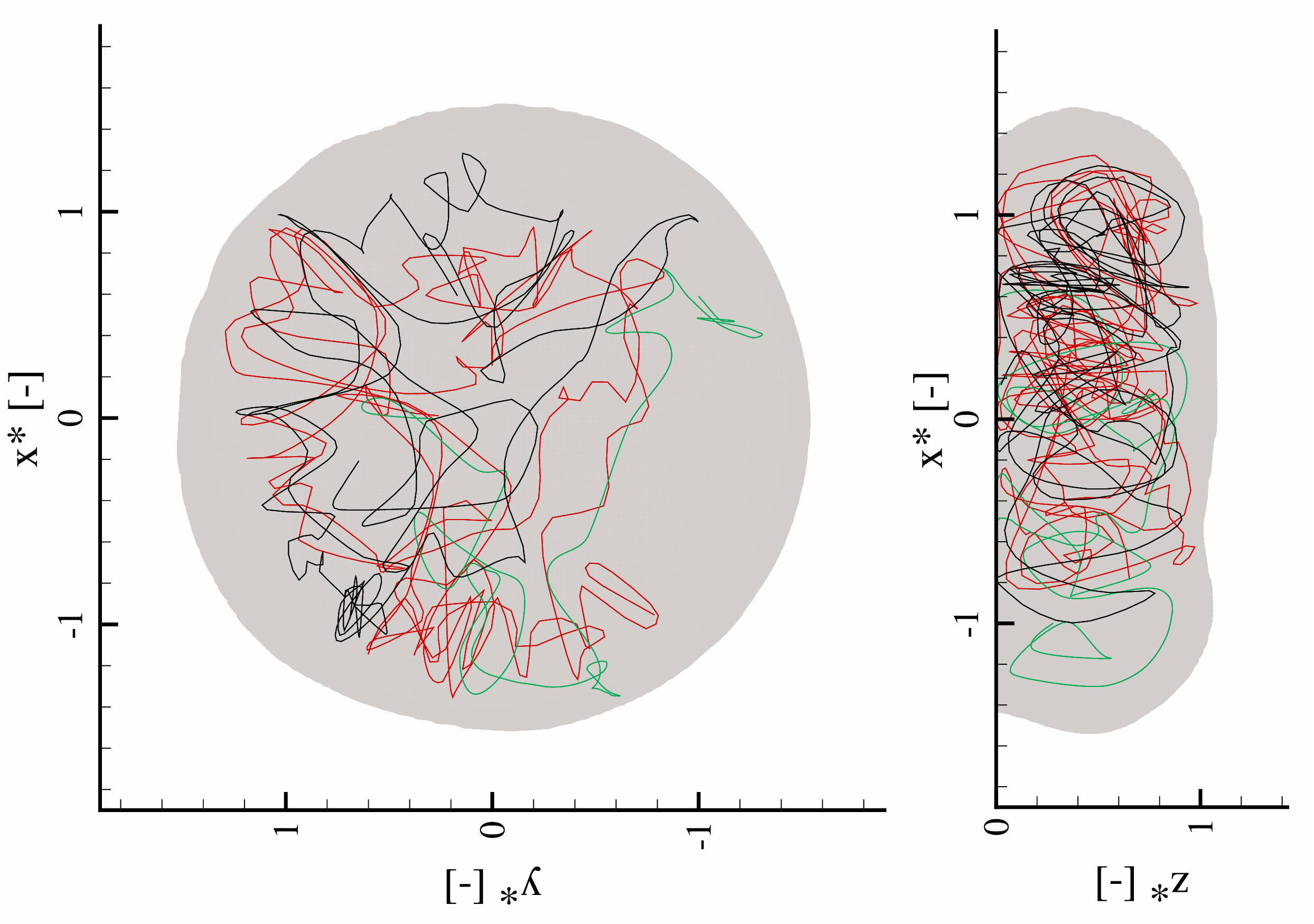}

\protect\caption{Particle tracks for cases 3 (left half) and 4 (right half), showing regular unsteady motion for case 3 and chaotic motion for case 4. \label{fig:Particle-tracks}}
\end{figure}

We have introduced three massless tracer particles into the pool in our simulations to obtain particle traces over a time of \SI{1}{\second} (Fig.~\ref{fig:Particle-tracks}), useful for comparison with experimental radiography results. For case 2 (not shown) and 3, the donut-shaped vortex-ring can be recognized fairly well in the particle tracks. It is remarkable that some of the particles traverse a very large area of the pool, whereas others remained local to a certain region. The tracks obtained agree qualitatively with experimental results by \citet{Czerner2005Schmelzbaddynamik}.
For cases 4 and 5 (latter not shown), all particles have moved through a large part of the domain, and the particle track projections are difficult to interpret, indicating turbulent flow. The particle track for case 1 is not shown as the laminar flow obtained leads to a very regular, linear pattern without deviations in the azimuthal direction.

\FloatBarrier

\subsection{Turbulence in the pool}

To quantify the level of turbulence in the pool, we placed a monitoring probe at  $(x^*,y^*,z^*)=(0, 0, 3/14)$ for cases 1-3, and $(x^*,y^*,z^*)=(15/14, 0, 3/14)$ for cases 4 and 5 (cf. Fig.~\ref{fig:3D-mesh}). The monitoring point is thus placed close to the free surface and in proximity to the downward jet emerging from the stagnation line at the free surface for all cases.

\begin{figure}
\subfloat[Case 2 and 3, point $M_{1,2,3}$] {
  \includegraphics[width=0.98\textwidth]{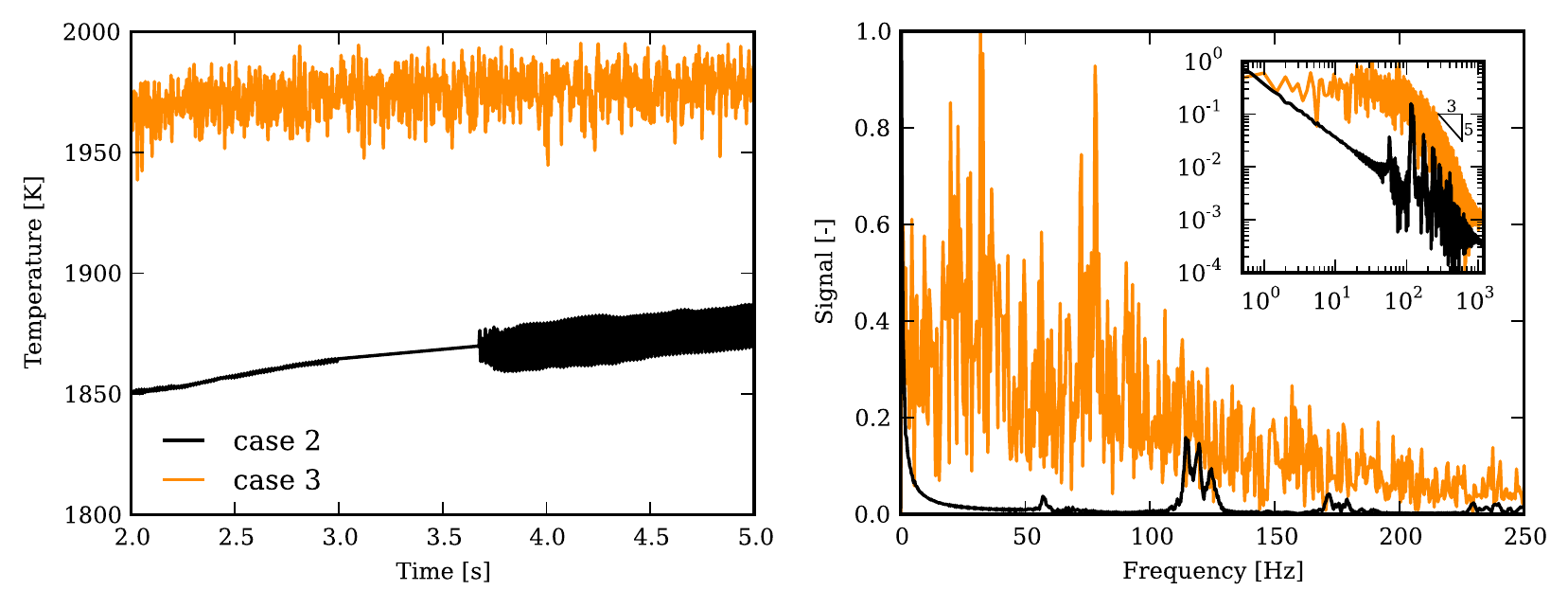}
\label{fig:Tprobe1519}
} \\
\subfloat[Case 4 and 5, point $M_{4,5}$]{
  \includegraphics[width=0.98\textwidth]{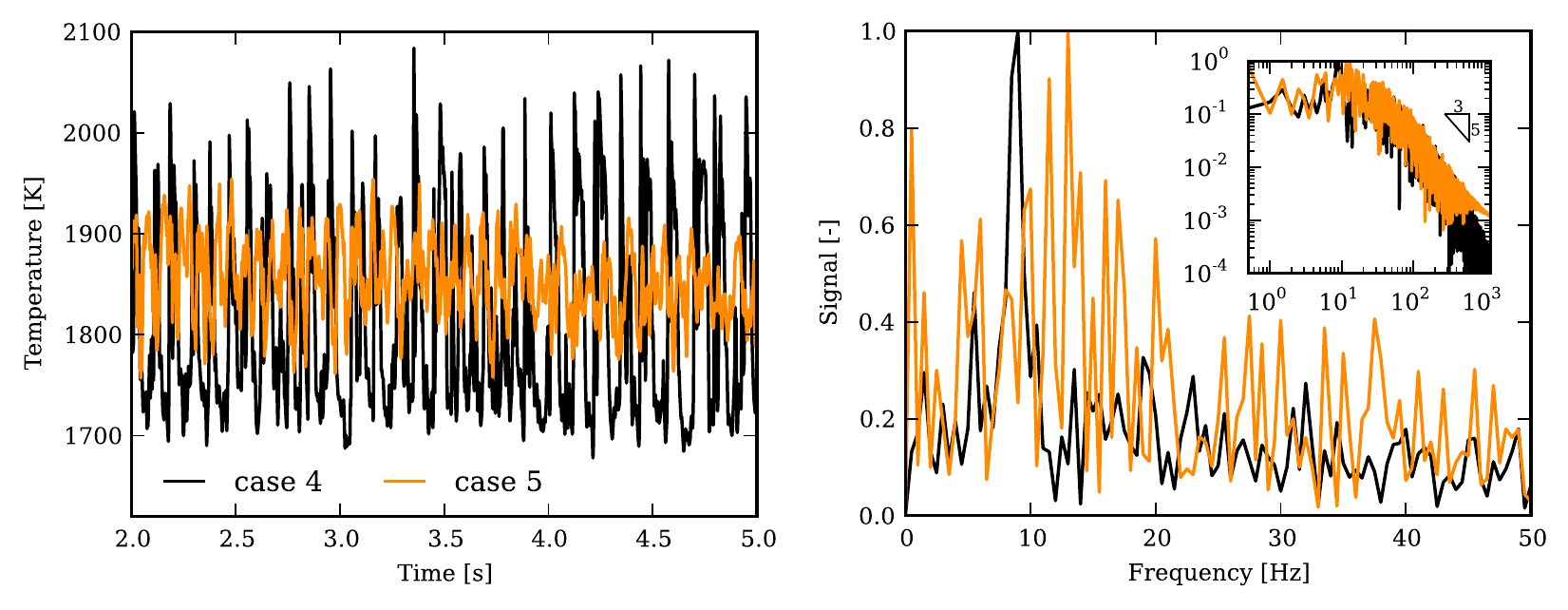}
\label{fig:Tprobe3852}
}
\protect\caption{Temperature signal and spectrum at monitoring points (see Fig.~\ref{fig:3D-mesh}) for varying laser powers. The signal spectrum has been normalized with the respective maximum peak.\label{fig:Tprobe}}
\end{figure}

The temperature history at these monitoring points is shown in Fig.~ \ref{fig:Tprobe}. After roughly \SI{2}{\second}, the temperature signal in all cases reaches a quasi steady state, where the conductive heat loss to the solid material more or less matches the heat input. For case 1, the temperature signal is smooth, again indicating laminar flow (not shown here). The onset of rotational motion in case 2 cannot be seen in the probe history at the pool center. Only when the pulsation begins, the temperature signal at this location begins to oscillate with low amplitude and high frequency, with the most significant peak at \SI{120}{\hertz}. The temperature signal recorded for case 3 also oscillates at high frequency with a rather low amplitude. The signal spectrum is fairly uniform up to \SI{100}{\hertz}, where it drops. At the two highest laser powers, cases 4 and 5, the oscillation frequency is lower but amplitude is significantly higher. The signal spectrum drops off much sooner at around \SI{10}{\hertz} for case 4, with a sharp peak at \SI{9}{\hertz}. For case 5, the maximum frequency is shifted to \SI{12.5}{\hertz} and the frequency drops after \SI{20}{\hertz}.

\begin{figure}
\includegraphics[width=0.99\textwidth]{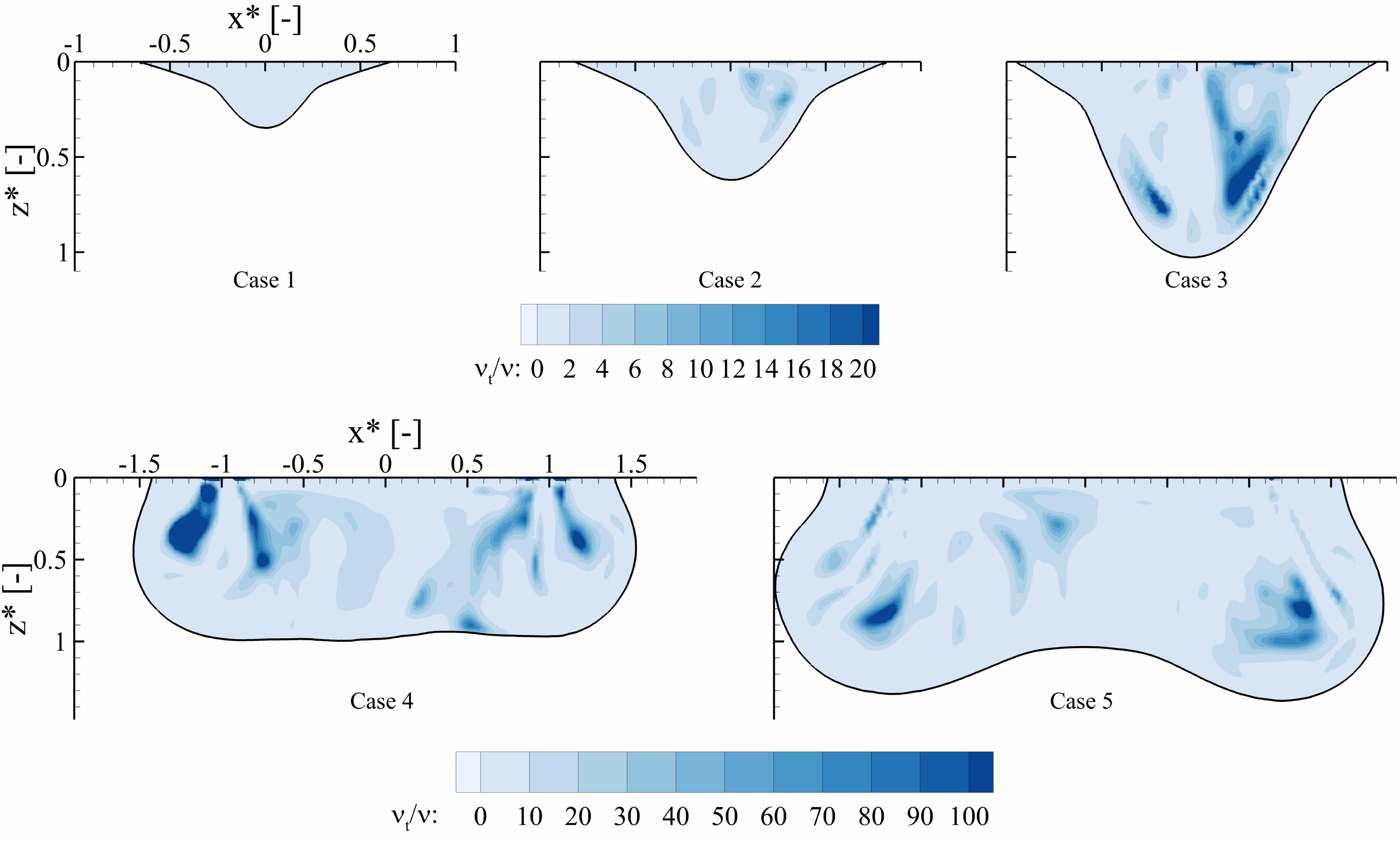}
\protect\caption{
Ratio of turbulent diffusivity over molecular diffusivity $\nu_{t}/\nu$
in the $y^*=0$ plane, based on turbulent kinetic energy
and turbulence dissipation averaged over the 4.5 to \SI{5}{\second} period.\label{fig:nu_turb_over_nu}
}

\end{figure}

Using the computed instantaneous velocity fluctuations $U'=U-\bar{U}$, we can determine the turbulent viscosity as $\nu_{t}=0.09k^{2}/\epsilon$ (Fig.~\ref{fig:nu_turb_over_nu}), with the turbulent kinetic energy $k=\overline{U'\cdot U'}/2$\nomenclature[ak]{$k$}{Turbulent kinetic energy} and the turbulent kinetic energy dissipation rate $\epsilon=\nu\overline{\nabla U':\nabla U'}$. Here, all averages, denoted by an overbar, have been computed over the time interval between 4.5 and 5\si{\second}. 
Unsurprisingly the turbulent viscosity is zero for the lowest power (case 1), since the flow is stable and laminar. For case 2, the turbulent viscosity is fairly uniform, with the space averaged value being of the same order of the molecular viscosity. 

For case 3, the local turbulent viscosity observed in the return flow of the jet from the pool bottom reaches values up to 50 times the molecular value, much higher than the values at the lower laser powers. The space averaged turbulent viscosity is 4 times the molecular value. For case 4, higher values of up to 400 times the molecular viscosity are obtained surrounding the jet stemming from the stagnation ring at the free surface. The space averaged value of the turbulent viscosity is 18 times the molecular value, and thus the contribution of momentum diffusivity due to turbulence is significant. A very similar distribution is obtained for the highest laser power, case 5. However, the maximum value of 200 times the molecular value is somewhat lower, and so is the space averaged value of 7 times the molecular value.

\begin{figure}
\includegraphics[width=1\textwidth]{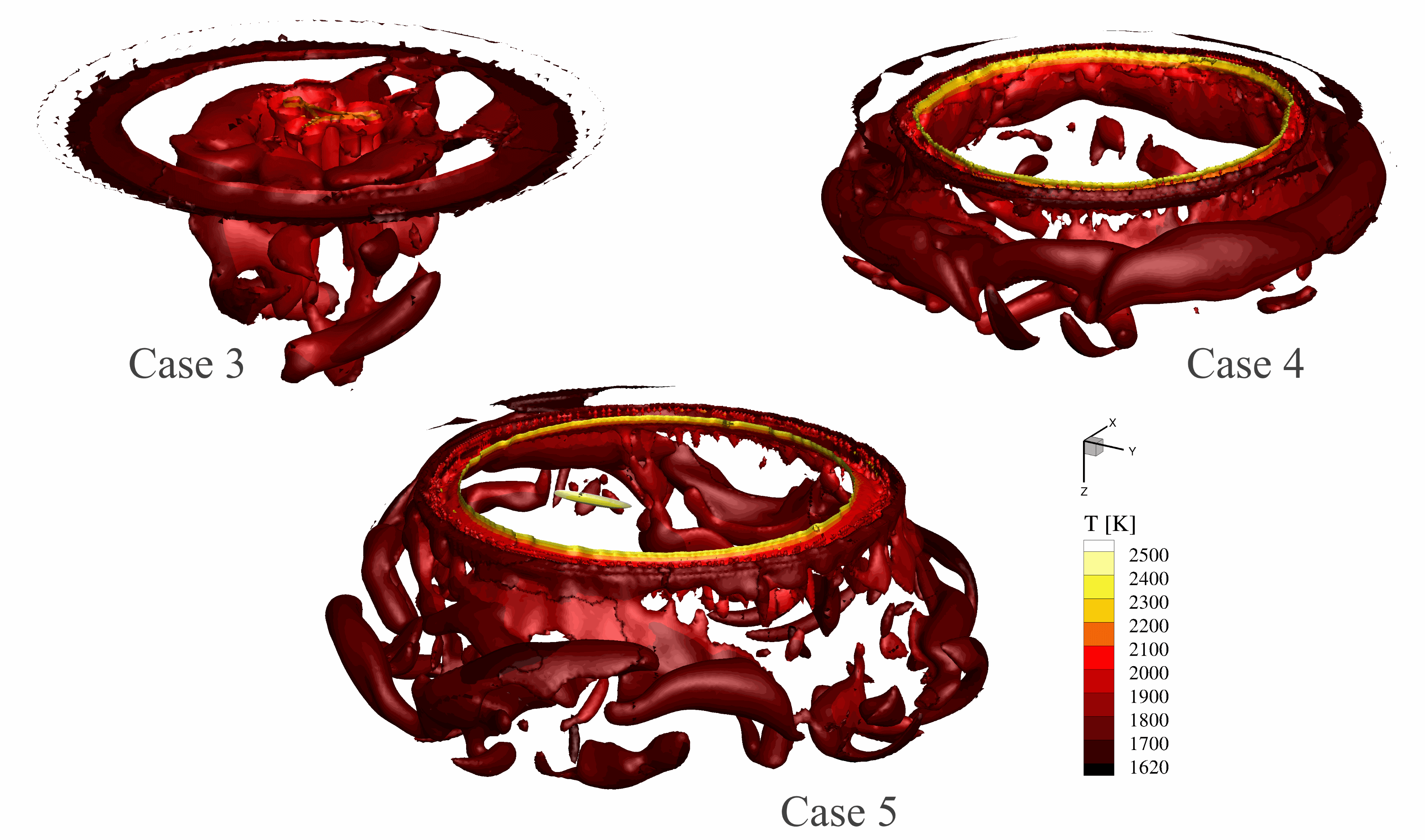}

\protect\caption{Isosurface of Q at a time instance, colored according to temperature, visualizing vortex tubes in the unsteady liquid pools for case 3, case 4 and case 5 (movies online).\label{fig:Qiso}}

\end{figure}

In Fig.~\ref{fig:Qiso} we show an isosurface of Q, the second invariant of $\nabla U$, \citep{Dubief2000Coherentvortex} to visualize coherent vortices in the pool. For case 3, we can identify a few coherent structures which span a large area of the pool, wrapping around in a corkscrew-like shape. Similarly to the temperature patterns at the free surface discussed earlier, the vortex structures rotate with time. For case 4, a large coherent structure appears, with a long connected vortex tube that spans the entire pool circumference and few isolated tubes around it. At the highest laser power, case 5, the structure has broken up significantly, and is much less uniform.

\section{Conclusions}

We studied the flow and heat transfer in low Prandtl number liquid pools driven by thermocapillary forces that arise due to surface tension gradients caused by temperature gradients across the free surface. Unlike previous studies on thermocapillary instabilities, we take into account phase change and the presence of a surface active species, which dramatically changes the relationship between surface tension $\gamma$ and temperature $T$, leading to a maximum in the surface tension at a temperature $T_c$. Qualitative experimental observations\citep{Zhao2009Unsteady,Zhao2009Complex,Zhao2010Effect,Karcher2000Turbulent,Czerner2005Schmelzbaddynamik,Azami2001Effect} of such systems have indicated the occurrence of flow instabilities and possibly even turbulence. 

We find a stable, laminar flow for a Marangoni number of \num{2.1e6}. A slightly higher Marangoni number of \num{2.8e6} initially triggers a self-sustained rotational instability, as long as the maximum temperature is below $T_c$. Once the critical temperature $T_c$ is exceeded and the surface tension temperature coefficient $\partial\gamma/\partial{}T$ locally changes its sign, a regular pulsating, non-turbulent motion is superposed on the rotation.

At Marangoni numbers of \num{4.6e6} and higher, temperatures so far above $T_c$ are reached that the Marangoni force directed towards the pool boundary has a clear effect on the free surface flow, noticeably decelerating or even reversing the flow, causing significant flow instabilities. The amplitude and frequency spectrum of temperature oscillations support the argument of enhanced heat and momentum transport due to turbulent flow in the melt pool. The computed turbulent viscosity is locally orders of magnitude larger than the molecular value, and averaged in space it is 4-20 times larger. This enhanced heat and momentum transport is in agreement with ad-hoc modifications necessary to match experimental results in previously published numerical studies of industrial processes involving thermocapillary flows in liquid low Prandtl number pools, where the flow has been assumed to be laminar. \citep{Pitscheneder1996Role,Anderson2010Origin,De2003Probing,De2004Smart,De2006Improving,Pavlyk2001Numerical,Tan2012Numerical}

It is noteworthy that the self-sustained oscillations arise even without taking into account complex interactions such as a deformable liquid-gas interface, temperature dependent material properties, non-uniform surfactant concentrations or additional, competing forces such as buoyancy.

\appendix

\begin{acknowledgments}
We would like to thank the European Commission for funding the MINTWELD
project (reference 229108) via the FP7-NMP program. We thank SURFsara
for the support in using the Lisa Compute Cluster (project MP-235-12).
\end{acknowledgments}

\section{Material properties}
The dimensioned material properties used are listed in table \ref{tab:Material-properties} for convenience. The respective material is a steel alloy (S705) with 150ppm of surfactant (sulfur).

\begin{table}[!h]
\protect\caption{Material properties\label{tab:Material-properties}}

\begin{ruledtabular}
{}%
\begin{tabular*}{15cm}{@{\extracolsep{\fill}}>{\raggedright}b{8cm}r>{\raggedright}b{2cm}}
{Property} & {Value} & {Unit}\tabularnewline
\midrule
{Solidus temperature $T_{s}$} & {$1610$} & {\si{\kelvin}}\tabularnewline
{Liquidus temperature $T_{l}$} & {$1620$} & {\si{\kelvin}}\tabularnewline
{Mushy zone porosity coefficient $\mu K_0$} & {$10^6$} & {\si{\newton\second\per\meter\tothe{4}}}\tabularnewline
{Specific heat capacity $c_{p}$} & {$670$} & {\si{\joule\per\kilo\gram\per\kelvin}}\tabularnewline
{Density $\rho$} & {$8100$} & {\si{\kilo\gram\per\cubic\meter}}\tabularnewline
{Thermal conductivity $\lambda$} & {$22.9$} & {\si{\watt\per\meter\per\kelvin}}\tabularnewline
{Latent heat of fusion $h_{f}$} & {$\num{2.508e5}$} & {\si{\joule\per\kilo\gram}}\tabularnewline
{Viscosity $\mu$} & {$\num{6e-3}$} & {\si{\pascal\second}}\tabularnewline
{Surface tension temperature coefficient $\partial\gamma/\partial T|_0$} & {$\num{-5.0e-4}$} & {\si{\newton\per\meter\per\kelvin}}\tabularnewline
{Entropy factor $k_l$} & {$\num{3.18e-3}$} & {$-$}\tabularnewline
{Standard heat of adsorption $\Delta H^0$} & {$\num{-1.66e8}$} & {\si{\joule\per\kilo\gram}}\tabularnewline
{Surface excess at saturation $\Gamma_s$} & {$\num{1.3e-8}$} & {\si{\kilo\mol\per\square\meter}}\tabularnewline
{Surfactant activity $a_s$} & {\num{150e-4}} & {wt\%}\tabularnewline
\end{tabular*}\end{ruledtabular}

\end{table}

\bibliography{kidess_pof}

\begin{thebibliography}{52}%
\makeatletter
\providecommand \@ifxundefined [1]{%
 \@ifx{#1\undefined}
}%
\providecommand \@ifnum [1]{%
 \ifnum #1\expandafter \@firstoftwo
 \else \expandafter \@secondoftwo
 \fi
}%
\providecommand \@ifx [1]{%
 \ifx #1\expandafter \@firstoftwo
 \else \expandafter \@secondoftwo
 \fi
}%
\providecommand \natexlab [1]{#1}%
\providecommand \enquote  [1]{``#1''}%
\providecommand \bibnamefont  [1]{#1}%
\providecommand \bibfnamefont [1]{#1}%
\providecommand \citenamefont [1]{#1}%
\providecommand \href@noop [0]{\@secondoftwo}%
\providecommand \href [0]{\begingroup \@sanitize@url \@href}%
\providecommand \@href[1]{\@@startlink{#1}\@@href}%
\providecommand \@@href[1]{\endgroup#1\@@endlink}%
\providecommand \@sanitize@url [0]{\catcode `\\12\catcode `\$12\catcode
  `\&12\catcode `\#12\catcode `\^12\catcode `\_12\catcode `\%12\relax}%
\providecommand \@@startlink[1]{}%
\providecommand \@@endlink[0]{}%
\providecommand \url  [0]{\begingroup\@sanitize@url \@url }%
\providecommand \@url [1]{\endgroup\@href {#1}{\urlprefix }}%
\providecommand \urlprefix  [0]{URL }%
\providecommand \Eprint [0]{\href }%
\providecommand \doibase [0]{http://dx.doi.org/}%
\providecommand \selectlanguage [0]{\@gobble}%
\providecommand \bibinfo  [0]{\@secondoftwo}%
\providecommand \bibfield  [0]{\@secondoftwo}%
\providecommand \translation [1]{[#1]}%
\providecommand \BibitemOpen [0]{}%
\providecommand \bibitemStop [0]{}%
\providecommand \bibitemNoStop [0]{.\EOS\space}%
\providecommand \EOS [0]{\spacefactor3000\relax}%
\providecommand \BibitemShut  [1]{\csname bibitem#1\endcsname}%
\let\auto@bib@innerbib\@empty
\bibitem [{\citenamefont {Schatz}\ and\ \citenamefont
  {Neitzel}(2001)}]{Schatz2001EXPERIMENTS}%
  \BibitemOpen
  \bibfield  {author} {\bibinfo {author} {\bibfnamefont {M.~F.}\ \bibnamefont
  {Schatz}}\ and\ \bibinfo {author} {\bibfnamefont {G.~P.}\ \bibnamefont
  {Neitzel}},\ }\bibfield  {title} {\enquote {\bibinfo {title} {Experiments on
  thermocapillary instabilities},}\ }\href {\doibase
  10.1146/annurev.fluid.33.1.93} {\bibfield  {journal} {\bibinfo  {journal}
  {Annual Review of Fluid Mechanics}\ }\textbf {\bibinfo {volume} {33}},\
  \bibinfo {pages} {93--127} (\bibinfo {year} {2001})}\BibitemShut {NoStop}%
\bibitem [{\citenamefont {Ohnishi}, \citenamefont {Azuma},\ and\ \citenamefont
  {Doi}(1992)}]{Ohnishi1992Computer}%
  \BibitemOpen
  \bibfield  {author} {\bibinfo {author} {\bibfnamefont {M.}~\bibnamefont
  {Ohnishi}}, \bibinfo {author} {\bibfnamefont {H.}~\bibnamefont {Azuma}}, \
  and\ \bibinfo {author} {\bibfnamefont {T.}~\bibnamefont {Doi}},\ }\bibfield
  {title} {\enquote {\bibinfo {title} {{Computer simulation of oscillatory
  Marangoni flow}},}\ }\href {\doibase 10.1016/0094-5765(92)90158-f} {\bibfield
   {journal} {\bibinfo  {journal} {Acta Astronautica}\ }\textbf {\bibinfo
  {volume} {26}},\ \bibinfo {pages} {685--696} (\bibinfo {year}
  {1992})}\BibitemShut {NoStop}%
\bibitem [{\citenamefont {Morvan}\ and\ \citenamefont
  {Bournot}(1996)}]{Morvan1996Oscillatory}%
  \BibitemOpen
  \bibfield  {author} {\bibinfo {author} {\bibfnamefont {D.}~\bibnamefont
  {Morvan}}\ and\ \bibinfo {author} {\bibfnamefont {P.}~\bibnamefont
  {Bournot}},\ }\bibfield  {title} {\enquote {\bibinfo {title} {{Oscillatory
  flow convection in a melted pool}},}\ }\href {\doibase
  10.1108/eum0000000004128} {\bibfield  {journal} {\bibinfo  {journal}
  {International Journal of Numerical Methods for Heat \& Fluid Flow}\ }\textbf
  {\bibinfo {volume} {6}},\ \bibinfo {pages} {13--20} (\bibinfo {year}
  {1996})}\BibitemShut {NoStop}%
\bibitem [{\citenamefont {Bucchignani}\ and\ \citenamefont
  {Mansutti}(2004)}]{Bucchignani2004RayleighMarangoni}%
  \BibitemOpen
  \bibfield  {author} {\bibinfo {author} {\bibfnamefont {E.}~\bibnamefont
  {Bucchignani}}\ and\ \bibinfo {author} {\bibfnamefont {D.}~\bibnamefont
  {Mansutti}},\ }\bibfield  {title} {\enquote {\bibinfo {title}
  {{Rayleigh-Marangoni} horizontal convection of low {Prandtl} number
  fluids},}\ }\href {\doibase 10.1063/1.1772381} {\bibfield  {journal}
  {\bibinfo  {journal} {Physics of Fluids (1994-present)}\ }\textbf {\bibinfo
  {volume} {16}},\ \bibinfo {pages} {3269--3280} (\bibinfo {year}
  {2004})}\BibitemShut {NoStop}%
\bibitem [{\citenamefont {Levenstam}, \citenamefont {Amberg},\ and\
  \citenamefont {Winkler}(2001)}]{Levenstam2001Instabilities}%
  \BibitemOpen
  \bibfield  {author} {\bibinfo {author} {\bibfnamefont {M.}~\bibnamefont
  {Levenstam}}, \bibinfo {author} {\bibfnamefont {G.}~\bibnamefont {Amberg}}, \
  and\ \bibinfo {author} {\bibfnamefont {C.}~\bibnamefont {Winkler}},\
  }\bibfield  {title} {\enquote {\bibinfo {title} {Instabilities of
  thermocapillary convection in a half-zone at intermediate {Prandtl}
  numbers},}\ }\href {\doibase 10.1063/1.1337063} {\bibfield  {journal}
  {\bibinfo  {journal} {Physics of Fluids (1994-present)}\ }\textbf {\bibinfo
  {volume} {13}},\ \bibinfo {pages} {807--816} (\bibinfo {year}
  {2001})}\BibitemShut {NoStop}%
\bibitem [{\citenamefont {Lappa}(2003)}]{Lappa2003Threedimensional}%
  \BibitemOpen
  \bibfield  {author} {\bibinfo {author} {\bibfnamefont {M.}~\bibnamefont
  {Lappa}},\ }\bibfield  {title} {\enquote {\bibinfo {title} {Three-dimensional
  numerical simulation of {Marangoni} flow instabilities in floating zones
  laterally heated by an equatorial ring},}\ }\href {\doibase
  10.1063/1.1543147} {\bibfield  {journal} {\bibinfo  {journal} {Physics of
  Fluids (1994-present)}\ }\textbf {\bibinfo {volume} {15}},\ \bibinfo {pages}
  {776--789} (\bibinfo {year} {2003})}\BibitemShut {NoStop}%
\bibitem [{\citenamefont {Kamotani}, \citenamefont {Matsumoto},\ and\
  \citenamefont {Yoda}(2007)}]{Kamotani2007Recent}%
  \BibitemOpen
  \bibfield  {author} {\bibinfo {author} {\bibfnamefont {Y.}~\bibnamefont
  {Kamotani}}, \bibinfo {author} {\bibfnamefont {S.}~\bibnamefont {Matsumoto}},
  \ and\ \bibinfo {author} {\bibfnamefont {S.}~\bibnamefont {Yoda}},\
  }\bibfield  {title} {\enquote {\bibinfo {title} {Recent developments in
  oscillatory {Marangoni} convection},}\ }\href {\doibase
  10.3970/fdmp.2007.003.147} {\bibfield  {journal} {\bibinfo  {journal} {Fluid
  Dynamics \& Materials Processing}\ }\textbf {\bibinfo {volume} {3}},\
  \bibinfo {pages} {147--160} (\bibinfo {year} {2007})}\BibitemShut {NoStop}%
\bibitem [{\citenamefont {Li}\ \emph {et~al.}(2004)\citenamefont {Li},
  \citenamefont {Imaishi}, \citenamefont {Azami},\ and\ \citenamefont
  {Hibiya}}]{Li2004Threedimensional}%
  \BibitemOpen
  \bibfield  {author} {\bibinfo {author} {\bibfnamefont {Y.-R.}\ \bibnamefont
  {Li}}, \bibinfo {author} {\bibfnamefont {N.}~\bibnamefont {Imaishi}},
  \bibinfo {author} {\bibfnamefont {T.}~\bibnamefont {Azami}}, \ and\ \bibinfo
  {author} {\bibfnamefont {T.}~\bibnamefont {Hibiya}},\ }\bibfield  {title}
  {\enquote {\bibinfo {title} {Three-dimensional oscillatory flow in a thin
  annular pool of silicon melt},}\ }\href {\doibase
  10.1016/j.jcrysgro.2003.08.017} {\bibfield  {journal} {\bibinfo  {journal}
  {Journal of Crystal Growth}\ }\textbf {\bibinfo {volume} {260}},\ \bibinfo
  {pages} {28--42} (\bibinfo {year} {2004})}\BibitemShut {NoStop}%
\bibitem [{\citenamefont {Xu}, \citenamefont {Ai},\ and\ \citenamefont
  {Li}(2007)}]{Xu2007RayleighBenardMarangoni}%
  \BibitemOpen
  \bibfield  {author} {\bibinfo {author} {\bibfnamefont {B.}~\bibnamefont
  {Xu}}, \bibinfo {author} {\bibfnamefont {X.}~\bibnamefont {Ai}}, \ and\
  \bibinfo {author} {\bibfnamefont {B.~Q.}\ \bibnamefont {Li}},\ }\bibfield
  {title} {\enquote {\bibinfo {title} {{Rayleigh-B\'{e}nard-Marangoni}
  instabilities of low-{Prandtl}-number fluid in a vertical cylinder with
  lateral heating},}\ }\href {\doibase 10.1080/10407780601009074} {\bibfield
  {journal} {\bibinfo  {journal} {Numerical Heat Transfer, Part A:
  Applications}\ }\textbf {\bibinfo {volume} {51}},\ \bibinfo {pages}
  {1119--1135} (\bibinfo {year} {2007})}\BibitemShut {NoStop}%
\bibitem [{\citenamefont {Sahoo}, \citenamefont {DebRoy},\ and\ \citenamefont
  {McNallan}(1988)}]{Sahoo1988Surface}%
  \BibitemOpen
  \bibfield  {author} {\bibinfo {author} {\bibfnamefont {P.}~\bibnamefont
  {Sahoo}}, \bibinfo {author} {\bibfnamefont {T.}~\bibnamefont {DebRoy}}, \
  and\ \bibinfo {author} {\bibfnamefont {M.}~\bibnamefont {McNallan}},\
  }\bibfield  {title} {\enquote {\bibinfo {title} {{Surface tension of binary
  metal - surface active solute systems under conditions relevant to welding
  metallurgy}},}\ }\href {\doibase 10.1007/bf02657748} {\bibfield  {journal}
  {\bibinfo  {journal} {Metallurgical and Materials Transactions B}\ }\textbf
  {\bibinfo {volume} {19}},\ \bibinfo {pages} {483--491--491} (\bibinfo {year}
  {1988})}\BibitemShut {NoStop}%
\bibitem [{\citenamefont {Hibiya}, \citenamefont {Morohoshi},\ and\
  \citenamefont {Ozawa}(2010)}]{Hibiya2010Oxygen}%
  \BibitemOpen
  \bibfield  {author} {\bibinfo {author} {\bibfnamefont {T.}~\bibnamefont
  {Hibiya}}, \bibinfo {author} {\bibfnamefont {K.}~\bibnamefont {Morohoshi}}, \
  and\ \bibinfo {author} {\bibfnamefont {S.}~\bibnamefont {Ozawa}},\ }\bibfield
   {title} {\enquote {\bibinfo {title} {Oxygen partial pressure dependence of
  surface tension and its temperature coefficient for metallic melts: a
  discussion from the viewpoint of solubility and adsorption of oxygen},}\
  }\href {\doibase 10.1007/s10853-009-4107-2} {\bibfield  {journal} {\bibinfo
  {journal} {Journal of Materials Science}\ }\textbf {\bibinfo {volume} {45}},\
  \bibinfo {pages} {1986--1992} (\bibinfo {year} {2010})}\BibitemShut {NoStop}%
\bibitem [{\citenamefont {Azami}, \citenamefont {Nakamura},\ and\ \citenamefont
  {Hibiya}(2001)}]{Azami2001Effect}%
  \BibitemOpen
  \bibfield  {author} {\bibinfo {author} {\bibfnamefont {T.}~\bibnamefont
  {Azami}}, \bibinfo {author} {\bibfnamefont {S.}~\bibnamefont {Nakamura}}, \
  and\ \bibinfo {author} {\bibfnamefont {T.}~\bibnamefont {Hibiya}},\
  }\bibfield  {title} {\enquote {\bibinfo {title} {Effect of oxygen on
  thermocapillary convection in a molten silicon column under microgravity},}\
  }\href {\doibase 10.1149/1.1353579} {\bibfield  {journal} {\bibinfo
  {journal} {Journal of The Electrochemical Society}\ }\textbf {\bibinfo
  {volume} {148}},\ \bibinfo {pages} {G185--G189} (\bibinfo {year}
  {2001})}\BibitemShut {NoStop}%
\bibitem [{\citenamefont {Zhao}\ \emph {et~al.}(2009)\citenamefont {Zhao},
  \citenamefont {van Steijn}, \citenamefont {Richardson}, \citenamefont
  {Kleijn}, \citenamefont {Kenjeres},\ and\ \citenamefont
  {Saldi}}]{Zhao2009Unsteady}%
  \BibitemOpen
  \bibfield  {author} {\bibinfo {author} {\bibfnamefont {C.~X.}\ \bibnamefont
  {Zhao}}, \bibinfo {author} {\bibfnamefont {V.}~\bibnamefont {van Steijn}},
  \bibinfo {author} {\bibfnamefont {I.~M.}\ \bibnamefont {Richardson}},
  \bibinfo {author} {\bibfnamefont {C.~R.}\ \bibnamefont {Kleijn}}, \bibinfo
  {author} {\bibfnamefont {S.}~\bibnamefont {Kenjeres}}, \ and\ \bibinfo
  {author} {\bibfnamefont {Z.}~\bibnamefont {Saldi}},\ }\bibfield  {title}
  {\enquote {\bibinfo {title} {{Unsteady interfacial phenomena during inward
  weld pool flow with an active surface oxide}},}\ }\href {\doibase
  10.1179/136217108x370281} {\bibfield  {journal} {\bibinfo  {journal} {Science
  and Technology of Welding \& Joining}\ }\textbf {\bibinfo {volume} {14}},\
  \bibinfo {pages} {132--140} (\bibinfo {year} {2009})}\BibitemShut {NoStop}%
\bibitem [{\citenamefont {Zhao}\ \emph {et~al.}(2010)\citenamefont {Zhao},
  \citenamefont {Kwakernaak}, \citenamefont {Pan}, \citenamefont {Richardson},
  \citenamefont {Saldi}, \citenamefont {Kenjeres},\ and\ \citenamefont
  {Kleijn}}]{Zhao2010Effect}%
  \BibitemOpen
  \bibfield  {author} {\bibinfo {author} {\bibfnamefont {C.~X.}\ \bibnamefont
  {Zhao}}, \bibinfo {author} {\bibfnamefont {C.}~\bibnamefont {Kwakernaak}},
  \bibinfo {author} {\bibfnamefont {Y.}~\bibnamefont {Pan}}, \bibinfo {author}
  {\bibfnamefont {I.~M.}\ \bibnamefont {Richardson}}, \bibinfo {author}
  {\bibfnamefont {Z.}~\bibnamefont {Saldi}}, \bibinfo {author} {\bibfnamefont
  {S.}~\bibnamefont {Kenjeres}}, \ and\ \bibinfo {author} {\bibfnamefont
  {C.~R.}\ \bibnamefont {Kleijn}},\ }\bibfield  {title} {\enquote {\bibinfo
  {title} {{The effect of oxygen on transitional Marangoni flow in laser spot
  welding}},}\ }\href {\doibase 10.1016/j.actamat.2010.07.056} {\bibfield
  {journal} {\bibinfo  {journal} {Acta Materialia}\ }\textbf {\bibinfo {volume}
  {58}},\ \bibinfo {pages} {6345--6357} (\bibinfo {year} {2010})}\BibitemShut
  {NoStop}%
\bibitem [{\citenamefont {Kou}, \citenamefont {Limmaneevichitr},\ and\
  \citenamefont {Wei}(2011)}]{kou2011oscillatory}%
  \BibitemOpen
  \bibfield  {author} {\bibinfo {author} {\bibfnamefont {S.}~\bibnamefont
  {Kou}}, \bibinfo {author} {\bibfnamefont {C.}~\bibnamefont
  {Limmaneevichitr}}, \ and\ \bibinfo {author} {\bibfnamefont {P.}~\bibnamefont
  {Wei}},\ }\bibfield  {title} {\enquote {\bibinfo {title} {Oscillatory
  {Marangoni} flow: A fundamental study by conduction-mode laser spot
  welding},}\ }\href
  {http://www.americanweldingsociety.org/wj/supplement/WJ\_201112\_s229.pdf}
  {\bibfield  {journal} {\bibinfo  {journal} {Welding Journal}\ }\textbf
  {\bibinfo {volume} {90}} (\bibinfo {year} {2011})}\BibitemShut {NoStop}%
\bibitem [{\citenamefont {Winkler}\ and\ \citenamefont
  {Amberg}(2005)}]{Winker2005Multicomponent}%
  \BibitemOpen
  \bibfield  {author} {\bibinfo {author} {\bibfnamefont {C.}~\bibnamefont
  {Winkler}}\ and\ \bibinfo {author} {\bibfnamefont {G.}~\bibnamefont
  {Amberg}},\ }\bibfield  {title} {\enquote {\bibinfo {title} {Multicomponent
  surfactant mass transfer in {GTA}-welding},}\ }\href {\doibase
  10.1504/pcfd.2005.006754} {\bibfield  {journal} {\bibinfo  {journal}
  {Progress in Computational Fluid Dynamics, An International Journal}\
  }\textbf {\bibinfo {volume} {5}},\ \bibinfo {pages} {190--206} (\bibinfo
  {year} {2005})}\BibitemShut {NoStop}%
\bibitem [{\citenamefont {Do-Quang}, \citenamefont {Amberg},\ and\
  \citenamefont {Pettersson}(2008)}]{DoQuang2008Modeling}%
  \BibitemOpen
  \bibfield  {author} {\bibinfo {author} {\bibfnamefont {M.}~\bibnamefont
  {Do-Quang}}, \bibinfo {author} {\bibfnamefont {G.}~\bibnamefont {Amberg}}, \
  and\ \bibinfo {author} {\bibfnamefont {C.-O.}\ \bibnamefont {Pettersson}},\
  }\bibfield  {title} {\enquote {\bibinfo {title} {Modeling of the adsorption
  kinetics and the convection of surfactants in a weld pool},}\ }\href
  {\doibase 10.1115/1.2946476} {\bibfield  {journal} {\bibinfo  {journal}
  {Journal of Heat Transfer}\ }\textbf {\bibinfo {volume} {130}},\ \bibinfo
  {pages} {092102+} (\bibinfo {year} {2008})}\BibitemShut {NoStop}%
\bibitem [{\citenamefont {Pumir}\ and\ \citenamefont
  {Blumenfeld}(1996)}]{Pumir1996Heat}%
  \BibitemOpen
  \bibfield  {author} {\bibinfo {author} {\bibfnamefont {A.}~\bibnamefont
  {Pumir}}\ and\ \bibinfo {author} {\bibfnamefont {L.}~\bibnamefont
  {Blumenfeld}},\ }\bibfield  {title} {\enquote {\bibinfo {title} {{Heat
  transport in a liquid layer locally heated on its free surface}},}\ }\href
  {\doibase 10.1103/physreve.54.r4528} {\bibfield  {journal} {\bibinfo
  {journal} {Physical Review E}\ }\textbf {\bibinfo {volume} {54}},\ \bibinfo
  {pages} {R4528--R4531} (\bibinfo {year} {1996})}\BibitemShut {NoStop}%
\bibitem [{\citenamefont {Karcher}\ \emph {et~al.}(2000)\citenamefont
  {Karcher}, \citenamefont {Schaller}, \citenamefont {Boeck}, \citenamefont
  {Metzner},\ and\ \citenamefont {Thess}}]{Karcher2000Turbulent}%
  \BibitemOpen
  \bibfield  {author} {\bibinfo {author} {\bibfnamefont {C.}~\bibnamefont
  {Karcher}}, \bibinfo {author} {\bibfnamefont {R.}~\bibnamefont {Schaller}},
  \bibinfo {author} {\bibfnamefont {T.}~\bibnamefont {Boeck}}, \bibinfo
  {author} {\bibfnamefont {C.}~\bibnamefont {Metzner}}, \ and\ \bibinfo
  {author} {\bibfnamefont {A.}~\bibnamefont {Thess}},\ }\bibfield  {title}
  {\enquote {\bibinfo {title} {{Turbulent heat transfer in liquid iron during
  electron beam evaporation}},}\ }\href {\doibase
  10.1016/s0017-9310(99)00248-3} {\bibfield  {journal} {\bibinfo  {journal}
  {International Journal of Heat and Mass Transfer}\ }\textbf {\bibinfo
  {volume} {43}},\ \bibinfo {pages} {1759--1766} (\bibinfo {year}
  {2000})}\BibitemShut {NoStop}%
\bibitem [{\citenamefont {Dikshit}\ \emph {et~al.}(2009)\citenamefont
  {Dikshit}, \citenamefont {Zende}, \citenamefont {Bhatia},\ and\ \citenamefont
  {Suri}}]{Dikshit2009Convection}%
  \BibitemOpen
  \bibfield  {author} {\bibinfo {author} {\bibfnamefont {B.}~\bibnamefont
  {Dikshit}}, \bibinfo {author} {\bibfnamefont {G.~R.}\ \bibnamefont {Zende}},
  \bibinfo {author} {\bibfnamefont {M.~S.}\ \bibnamefont {Bhatia}}, \ and\
  \bibinfo {author} {\bibfnamefont {B.~M.}\ \bibnamefont {Suri}},\ }\bibfield
  {title} {\enquote {\bibinfo {title} {Convection in molten pool created by a
  concentrated energy flux on a solid metal target},}\ }\href {\doibase
  10.1063/1.3210763} {\bibfield  {journal} {\bibinfo  {journal} {Physics of
  Fluids (1994-present)}\ }\textbf {\bibinfo {volume} {21}},\ \bibinfo {pages}
  {084105+} (\bibinfo {year} {2009})}\BibitemShut {NoStop}%
\bibitem [{\citenamefont {Boeck}\ and\ \citenamefont
  {Karcher}(2003)}]{Boeck2003LowPrandtlNumber}%
  \BibitemOpen
  \bibfield  {author} {\bibinfo {author} {\bibfnamefont {T.}~\bibnamefont
  {Boeck}}\ and\ \bibinfo {author} {\bibfnamefont {C.}~\bibnamefont
  {Karcher}},\ }\bibfield  {title} {\enquote {\bibinfo {title}
  {Low-{Prandtl}-number {Marangoni} convection driven by localized heating on
  the free surface: Results of three-dimensional direct simulations},}\ }in\
  \href {\doibase 10.1007/978-3-540-45095-5\_8} {\emph {\bibinfo {booktitle}
  {Interfacial Fluid Dynamics and Transport Processes}}},\ \bibinfo {series}
  {Lecture Notes in Physics}, Vol.\ \bibinfo {volume} {628},\ \bibinfo {editor}
  {edited by\ \bibinfo {editor} {\bibfnamefont {R.}~\bibnamefont {Narayanan}}\
  and\ \bibinfo {editor} {\bibfnamefont {D.}~\bibnamefont {Schwabe}}}\
  (\bibinfo  {publisher} {Springer Berlin Heidelberg},\ \bibinfo {year}
  {2003})\ pp.\ \bibinfo {pages} {157--175}\BibitemShut {NoStop}%
\bibitem [{\citenamefont {Kuhlmann}\ and\ \citenamefont
  {Schoisswohl}(2010)}]{Kuhlmann2010Flow}%
  \BibitemOpen
  \bibfield  {author} {\bibinfo {author} {\bibfnamefont {H.~C.}\ \bibnamefont
  {Kuhlmann}}\ and\ \bibinfo {author} {\bibfnamefont {U.}~\bibnamefont
  {Schoisswohl}},\ }\bibfield  {title} {\enquote {\bibinfo {title} {{Flow
  instabilities in thermocapillary-buoyant liquid pools}},}\ }\href {\doibase
  10.1017/s0022112009992953} {\bibfield  {journal} {\bibinfo  {journal}
  {Journal of Fluid Mechanics}\ }\textbf {\bibinfo {volume} {644}},\ \bibinfo
  {pages} {509--535} (\bibinfo {year} {2010})}\BibitemShut {NoStop}%
\bibitem [{\citenamefont {Singh}, \citenamefont {Pardeshi},\ and\ \citenamefont
  {Basu}(2001)}]{Singh2001Modelling}%
  \BibitemOpen
  \bibfield  {author} {\bibinfo {author} {\bibfnamefont {A.}~\bibnamefont
  {Singh}}, \bibinfo {author} {\bibfnamefont {R.}~\bibnamefont {Pardeshi}}, \
  and\ \bibinfo {author} {\bibfnamefont {B.}~\bibnamefont {Basu}},\ }\bibfield
  {title} {\enquote {\bibinfo {title} {{Modelling of convection during
  solidification of metal and alloys}},}\ }\href {\doibase 10.1007/bf02728483}
  {\bibfield  {journal} {\bibinfo  {journal} {Sadhana}\ }\textbf {\bibinfo
  {volume} {26}},\ \bibinfo {pages} {139--162} (\bibinfo {year}
  {2001})}\BibitemShut {NoStop}%
\bibitem [{\citenamefont {Pitscheneder}\ \emph {et~al.}(1996)\citenamefont
  {Pitscheneder}, \citenamefont {DebRoy}, \citenamefont {Mundra},\ and\
  \citenamefont {Ebner}}]{Pitscheneder1996Role}%
  \BibitemOpen
  \bibfield  {author} {\bibinfo {author} {\bibfnamefont {W.}~\bibnamefont
  {Pitscheneder}}, \bibinfo {author} {\bibfnamefont {T.}~\bibnamefont
  {DebRoy}}, \bibinfo {author} {\bibfnamefont {K.}~\bibnamefont {Mundra}}, \
  and\ \bibinfo {author} {\bibfnamefont {R.}~\bibnamefont {Ebner}},\ }\bibfield
   {title} {\enquote {\bibinfo {title} {Role of sulfur and processing variables
  on the temporal evolution of weld pool geometry during multikilowatt laser
  beam welding of steels},}\ }\href
  {http://www2.matse.psu.edu/modeling/short\_course/sulfur.pdf} {\bibfield
  {journal} {\bibinfo  {journal} {Welding Journal}\ }\textbf {\bibinfo {volume}
  {75}},\ \bibinfo {pages} {71--s--80--s} (\bibinfo {year} {1996})}\BibitemShut
  {NoStop}%
\bibitem [{\citenamefont {Hibiya}\ and\ \citenamefont
  {Ozawa}(2013)}]{Hibiya2013Effect}%
  \BibitemOpen
  \bibfield  {author} {\bibinfo {author} {\bibfnamefont {T.}~\bibnamefont
  {Hibiya}}\ and\ \bibinfo {author} {\bibfnamefont {S.}~\bibnamefont {Ozawa}},\
  }\bibfield  {title} {\enquote {\bibinfo {title} {Effect of oxygen partial
  pressure on the marangoni flow of molten metals},}\ }\href {\doibase
  10.1002/crat.201200514} {\bibfield  {journal} {\bibinfo  {journal} {Cryst.
  Res. Technol.}\ }\textbf {\bibinfo {volume} {48}},\ \bibinfo {pages}
  {208--213} (\bibinfo {year} {2013})}\BibitemShut {NoStop}%
\bibitem [{\citenamefont {Ozawa}\ \emph {et~al.}(2014)\citenamefont {Ozawa},
  \citenamefont {Takahashi}, \citenamefont {Watanabe},\ and\ \citenamefont
  {Fukuyama}}]{Ozawa2014Influence}%
  \BibitemOpen
  \bibfield  {author} {\bibinfo {author} {\bibfnamefont {S.}~\bibnamefont
  {Ozawa}}, \bibinfo {author} {\bibfnamefont {S.}~\bibnamefont {Takahashi}},
  \bibinfo {author} {\bibfnamefont {N.}~\bibnamefont {Watanabe}}, \ and\
  \bibinfo {author} {\bibfnamefont {H.}~\bibnamefont {Fukuyama}},\ }\bibfield
  {title} {\enquote {\bibinfo {title} {Influence of oxygen adsorption on
  surface tension of molten nickel measured under reducing gas atmosphere},}\
  }\bibfield  {booktitle} {\emph {\bibinfo {booktitle} {International Journal
  of Thermophysics}},\ }\href {\doibase 10.1007/s10765-014-1674-5} {\ \textbf
  {\bibinfo {volume} {35}},\ \bibinfo {pages} {1705--1711} (\bibinfo {year}
  {2014})}\BibitemShut {NoStop}%
\bibitem [{\citenamefont {Weller}\ \emph {et~al.}(1998)\citenamefont {Weller},
  \citenamefont {Tabor}, \citenamefont {Jasak},\ and\ \citenamefont
  {Fureby}}]{Weller1998Tensorial}%
  \BibitemOpen
  \bibfield  {author} {\bibinfo {author} {\bibfnamefont {H.~G.}\ \bibnamefont
  {Weller}}, \bibinfo {author} {\bibfnamefont {G.}~\bibnamefont {Tabor}},
  \bibinfo {author} {\bibfnamefont {H.}~\bibnamefont {Jasak}}, \ and\ \bibinfo
  {author} {\bibfnamefont {C.}~\bibnamefont {Fureby}},\ }\bibfield  {title}
  {\enquote {\bibinfo {title} {{A tensorial approach to computational continuum
  mechanics using object-oriented techniques}},}\ }\href {\doibase
  10.1063/1.168744} {\bibfield  {journal} {\bibinfo  {journal} {Computers in
  Physics}\ }\textbf {\bibinfo {volume} {12}},\ \bibinfo {pages} {620--631}
  (\bibinfo {year} {1998})}\BibitemShut {NoStop}%
\bibitem [{\citenamefont {Berberovic}(2010)}]{Berberovic2010Investigation}%
  \BibitemOpen
  \bibfield  {author} {\bibinfo {author} {\bibfnamefont {E.}~\bibnamefont
  {Berberovic}},\ }\emph {\bibinfo {title} {{Investigation of Free-surface Flow
  Associated with Drop Impact: Numerical Simulations and Theoretical
  Modeling}}},\ \href@noop {} {Ph.D. thesis},\ \bibinfo  {school} {Technische
  Universit{ae}t Darmstadt} (\bibinfo {year} {2010})\BibitemShut {NoStop}%
\bibitem [{\citenamefont {Issa}(1986)}]{Issa1986Solution}%
  \BibitemOpen
  \bibfield  {author} {\bibinfo {author} {\bibfnamefont {R.~I.}\ \bibnamefont
  {Issa}},\ }\bibfield  {title} {\enquote {\bibinfo {title} {{Solution of the
  implicitly discretised fluid flow equations by operator-splitting}},}\ }\href
  {\doibase 10.1016/0021-9991(86)90099-9} {\bibfield  {journal} {\bibinfo
  {journal} {Journal of Computational Physics}\ }\textbf {\bibinfo {volume}
  {62}},\ \bibinfo {pages} {40--65} (\bibinfo {year} {1986})}\BibitemShut
  {NoStop}%
\bibitem [{\citenamefont {Voller}\ and\ \citenamefont
  {Swaminathan}(1991)}]{Voller1991GENERAL}%
  \BibitemOpen
  \bibfield  {author} {\bibinfo {author} {\bibfnamefont {V.~R.}\ \bibnamefont
  {Voller}}\ and\ \bibinfo {author} {\bibfnamefont {C.~R.}\ \bibnamefont
  {Swaminathan}},\ }\bibfield  {title} {\enquote {\bibinfo {title} {General
  source-based method for solidification phase change},}\ }\href {\doibase
  10.1080/10407799108944962} {\bibfield  {journal} {\bibinfo  {journal}
  {Numerical Heat Transfer, Part B: Fundamentals: An International Journal of
  Computation and Methodology}\ }\textbf {\bibinfo {volume} {19}},\ \bibinfo
  {pages} {175--189} (\bibinfo {year} {1991})}\BibitemShut {NoStop}%
\bibitem [{Note1()}]{Note1}%
  \BibitemOpen
  \bibinfo {note} {\protect \url
  {https://surfsara.nl/systems/lisa}}\BibitemShut {NoStop}%
\bibitem [{\citenamefont {Lilly}(1992)}]{Lilly1992Proposed}%
  \BibitemOpen
  \bibfield  {author} {\bibinfo {author} {\bibfnamefont {D.~K.}\ \bibnamefont
  {Lilly}},\ }\bibfield  {title} {\enquote {\bibinfo {title} {A proposed
  modification of the {Germano} subgrid-scale closure method},}\ }\href
  {\doibase 10.1063/1.858280} {\bibfield  {journal} {\bibinfo  {journal}
  {Physics of Fluids A: Fluid Dynamics (1989-1993)}\ }\textbf {\bibinfo
  {volume} {4}},\ \bibinfo {pages} {633--635} (\bibinfo {year}
  {1992})}\BibitemShut {NoStop}%
\bibitem [{\citenamefont {Passalacqua}()}]{PassalacquaDynamicSmagorinsky}%
  \BibitemOpen
  \bibfield  {author} {\bibinfo {author} {\bibfnamefont {A.}~\bibnamefont
  {Passalacqua}},\ }\href {https://bitbucket.org/albertop/dynamicsmagorinsky}
  {\enquote {\bibinfo {title} {{dynamicSmagorinsky - Implementation of the
  dynamic Smagorinsky SGS model as proposed by Lilly (1992) for OpenFOAM}},}\
  }\bibinfo {howpublished} {Published electronically online,
  \url{https://bitbucket.org/albertop/dynamicsmagorinsky}}\BibitemShut
  {NoStop}%
\bibitem [{\citenamefont {Eidson}(1985)}]{Eidson1985Numerical}%
  \BibitemOpen
  \bibfield  {author} {\bibinfo {author} {\bibfnamefont {T.~M.}\ \bibnamefont
  {Eidson}},\ }\bibfield  {title} {\enquote {\bibinfo {title} {Numerical
  simulation of the turbulent {Rayleigh-B\'{e}nard} problem using subgrid
  modelling},}\ }\href {\doibase 10.1017/s0022112085002634} {\bibfield
  {journal} {\bibinfo  {journal} {Journal of Fluid Mechanics}\ }\textbf
  {\bibinfo {volume} {158}},\ \bibinfo {pages} {245--268} (\bibinfo {year}
  {1985})}\BibitemShut {NoStop}%
\bibitem [{\citenamefont {Kenjere\v{s}}\ and\ \citenamefont
  {Hanjali\'{c}}(2006)}]{Kenjeres2006LES}%
  \BibitemOpen
  \bibfield  {author} {\bibinfo {author} {\bibfnamefont {S.}~\bibnamefont
  {Kenjere\v{s}}}\ and\ \bibinfo {author} {\bibfnamefont {K.}~\bibnamefont
  {Hanjali\'{c}}},\ }\bibfield  {title} {\enquote {\bibinfo {title} {{LES,
  T-RANS and hybrid simulations of thermal convection at high Ra numbers}},}\
  }\href {\doibase 10.1016/j.ijheatfluidflow.2006.03.008} {\bibfield  {journal}
  {\bibinfo  {journal} {International Journal of Heat and Fluid Flow}\ }\textbf
  {\bibinfo {volume} {27}},\ \bibinfo {pages} {800--810} (\bibinfo {year}
  {2006})}\BibitemShut {NoStop}%
\bibitem [{\citenamefont {Zhao}\ and\ \citenamefont
  {Richardson}(2009)}]{Zhao2009Complex}%
  \BibitemOpen
  \bibfield  {author} {\bibinfo {author} {\bibfnamefont {C.}~\bibnamefont
  {Zhao}}\ and\ \bibinfo {author} {\bibfnamefont {I.}~\bibnamefont
  {Richardson}},\ }\bibfield  {title} {\enquote {\bibinfo {title} {Complex flow
  motions during laser welding},}\ }in\ \href {\doibase 10.2514/6.2009-3739}
  {\emph {\bibinfo {booktitle} {40th AIAA Plasmadynamics and Lasers
  Conference}}},\ \bibinfo {series and number} {Fluid Dynamics and Co-located
  Conferences}\ (\bibinfo  {publisher} {American Institute of Aeronautics and
  Astronautics},\ \bibinfo {year} {2009})\BibitemShut {NoStop}%
\bibitem [{\citenamefont {Smith}\ and\ \citenamefont
  {Davis}(1983)}]{SmithDavis1983Instabilities}%
  \BibitemOpen
  \bibfield  {author} {\bibinfo {author} {\bibfnamefont {M.~K.}\ \bibnamefont
  {Smith}}\ and\ \bibinfo {author} {\bibfnamefont {S.~H.}\ \bibnamefont
  {Davis}},\ }\bibfield  {title} {\enquote {\bibinfo {title} {Instabilities of
  dynamic thermocapillary liquid layers. part 1. convective instabilities},}\
  }\href {\doibase 10.1017/s0022112083001512} {\bibfield  {journal} {\bibinfo
  {journal} {Journal of Fluid Mechanics}\ }\textbf {\bibinfo {volume} {132}},\
  \bibinfo {pages} {119--144} (\bibinfo {year} {1983})}\BibitemShut {NoStop}%
\bibitem [{\citenamefont {Davis}(1987)}]{Davis1987Thermocapillary}%
  \BibitemOpen
  \bibfield  {author} {\bibinfo {author} {\bibfnamefont {S.~H.}\ \bibnamefont
  {Davis}},\ }\bibfield  {title} {\enquote {\bibinfo {title} {Thermocapillary
  instabilities},}\ }\href {\doibase 10.1146/annurev.fl.19.010187.002155}
  {\bibfield  {journal} {\bibinfo  {journal} {Annual Review of Fluid
  Mechanics}\ }\textbf {\bibinfo {volume} {19}},\ \bibinfo {pages} {403--435}
  (\bibinfo {year} {1987})}\BibitemShut {NoStop}%
\bibitem [{\citenamefont {Lappa}(2010)}]{Lappa2010Thermal}%
  \BibitemOpen
  \bibfield  {author} {\bibinfo {author} {\bibfnamefont {M.}~\bibnamefont
  {Lappa}},\ }\href@noop {} {\emph {\bibinfo {title} {Thermal convection
  patterns, evolution and stability}}}\ (\bibinfo  {publisher} {Wiley},\
  \bibinfo {year} {2010})\BibitemShut {NoStop}%
\bibitem [{\citenamefont {Li}\ \emph {et~al.}(2007)\citenamefont {Li},
  \citenamefont {Peng}, \citenamefont {Wu},\ and\ \citenamefont
  {Imaishi}}]{LiImaishi2007Bifurcation}%
  \BibitemOpen
  \bibfield  {author} {\bibinfo {author} {\bibfnamefont {Y.}~\bibnamefont
  {Li}}, \bibinfo {author} {\bibfnamefont {L.}~\bibnamefont {Peng}}, \bibinfo
  {author} {\bibfnamefont {S.}~\bibnamefont {Wu}}, \ and\ \bibinfo {author}
  {\bibfnamefont {N.}~\bibnamefont {Imaishi}},\ }\bibfield  {title} {\enquote
  {\bibinfo {title} {Bifurcation of thermocapillary convection in a shallow
  annular pool of silicon melt},}\ }\href {\doibase 10.1007/s10409-006-0053-2}
  {\bibfield  {journal} {\bibinfo  {journal} {Acta Mechanica Sinica}\ }\textbf
  {\bibinfo {volume} {23}},\ \bibinfo {pages} {43--48} (\bibinfo {year}
  {2007})}\BibitemShut {NoStop}%
\bibitem [{\citenamefont {Kidess}\ \emph {et~al.}(2016)\citenamefont {Kidess},
  \citenamefont {Kenjere\v{s}}, \citenamefont {Righolt},\ and\ \citenamefont
  {Kleijn}}]{Kidess2016Marangoni}%
  \BibitemOpen
  \bibfield  {author} {\bibinfo {author} {\bibfnamefont {A.}~\bibnamefont
  {Kidess}}, \bibinfo {author} {\bibfnamefont {S.}~\bibnamefont
  {Kenjere\v{s}}}, \bibinfo {author} {\bibfnamefont {B.~W.}\ \bibnamefont
  {Righolt}}, \ and\ \bibinfo {author} {\bibfnamefont {C.~R.}\ \bibnamefont
  {Kleijn}},\ }\bibfield  {title} {\enquote {\bibinfo {title} {Marangoni driven
  turbulence in high energy surface melting processes},}\ }\href {\doibase
  10.1016/j.ijthermalsci.2016.01.015} {\bibfield  {journal} {\bibinfo
  {journal} {International Journal of Thermal Sciences}\ } (\bibinfo {year}
  {2016}),\ 10.1016/j.ijthermalsci.2016.01.015},\ \bibinfo {note} {to
  appear}\BibitemShut {NoStop}%
\bibitem [{\citenamefont {Arata}\ \emph {et~al.}(1987)\citenamefont {Arata},
  \citenamefont {Tomie}, \citenamefont {Abe},\ and\ \citenamefont
  {Yao}}]{Arata1987}%
  \BibitemOpen
  \bibfield  {author} {\bibinfo {author} {\bibfnamefont {Y.}~\bibnamefont
  {Arata}}, \bibinfo {author} {\bibfnamefont {M.}~\bibnamefont {Tomie}},
  \bibinfo {author} {\bibfnamefont {N.}~\bibnamefont {Abe}}, \ and\ \bibinfo
  {author} {\bibfnamefont {X.-Y.}\ \bibnamefont {Yao}},\ }\bibfield  {title}
  {\enquote {\bibinfo {title} {Observation of molten metal flow during eb
  welding},}\ }\href {http://ci.nii.ac.jp/naid/110006486708/en/} {\bibfield
  {journal} {\bibinfo  {journal} {Transactions of JWRI}\ }\textbf {\bibinfo
  {volume} {16}},\ \bibinfo {pages} {13--16} (\bibinfo {year}
  {1987})}\BibitemShut {NoStop}%
\bibitem [{\citenamefont {Czerner}(2005)}]{Czerner2005Schmelzbaddynamik}%
  \BibitemOpen
  \bibfield  {author} {\bibinfo {author} {\bibfnamefont {S.}~\bibnamefont
  {Czerner}},\ }\emph {\bibinfo {title} {Schmelzbaddynamik beim
  Laserstrahl-W\"{a}rmeleitungsschwei{\ss}en von Eisenwerkstoffen}},\
  \href@noop {} {Ph.D. thesis},\ \bibinfo  {school} {Universit\"{a}t Hannover}
  (\bibinfo {year} {2005})\BibitemShut {NoStop}%
\bibitem [{\citenamefont {Gatzen}\ \emph {et~al.}(2011)\citenamefont {Gatzen},
  \citenamefont {Tang}, \citenamefont {Vollertsen}, \citenamefont {Mizutani},\
  and\ \citenamefont {Katayama}}]{Gatzen2011Xray}%
  \BibitemOpen
  \bibfield  {author} {\bibinfo {author} {\bibfnamefont {M.}~\bibnamefont
  {Gatzen}}, \bibinfo {author} {\bibfnamefont {Z.}~\bibnamefont {Tang}},
  \bibinfo {author} {\bibfnamefont {F.}~\bibnamefont {Vollertsen}}, \bibinfo
  {author} {\bibfnamefont {M.}~\bibnamefont {Mizutani}}, \ and\ \bibinfo
  {author} {\bibfnamefont {S.}~\bibnamefont {Katayama}},\ }\bibfield  {title}
  {\enquote {\bibinfo {title} {X-ray investigation of melt flow behavior under
  magnetic stirring regime in laser beam welding of aluminum},}\ }\href
  {\doibase 10.2351/1.3580552} {\bibfield  {journal} {\bibinfo  {journal}
  {Journal of Laser Applications}\ }\textbf {\bibinfo {volume} {23}},\ \bibinfo
  {pages} {032002+} (\bibinfo {year} {2011})}\BibitemShut {NoStop}%
\bibitem [{\citenamefont {Morisada}\ \emph {et~al.}(2011)\citenamefont
  {Morisada}, \citenamefont {Fujii}, \citenamefont {Kawahito}, \citenamefont
  {Nakata},\ and\ \citenamefont {Tanaka}}]{Morisada2011Threedimensional}%
  \BibitemOpen
  \bibfield  {author} {\bibinfo {author} {\bibfnamefont {Y.}~\bibnamefont
  {Morisada}}, \bibinfo {author} {\bibfnamefont {H.}~\bibnamefont {Fujii}},
  \bibinfo {author} {\bibfnamefont {Y.}~\bibnamefont {Kawahito}}, \bibinfo
  {author} {\bibfnamefont {K.}~\bibnamefont {Nakata}}, \ and\ \bibinfo {author}
  {\bibfnamefont {M.}~\bibnamefont {Tanaka}},\ }\bibfield  {title} {\enquote
  {\bibinfo {title} {Three-dimensional visualization of material flow during
  friction stir welding by two pairs of x-ray transmission systems},}\ }\href
  {\doibase 10.1016/j.scriptamat.2011.09.021} {\bibfield  {journal} {\bibinfo
  {journal} {Scripta Materialia}\ }\textbf {\bibinfo {volume} {65}},\ \bibinfo
  {pages} {1085--1088} (\bibinfo {year} {2011})}\BibitemShut {NoStop}%
\bibitem [{\citenamefont {Dubief}\ and\ \citenamefont
  {Delcayre}(2000)}]{Dubief2000Coherentvortex}%
  \BibitemOpen
  \bibfield  {author} {\bibinfo {author} {\bibfnamefont {Y.}~\bibnamefont
  {Dubief}}\ and\ \bibinfo {author} {\bibfnamefont {F.}~\bibnamefont
  {Delcayre}},\ }\bibfield  {title} {\enquote {\bibinfo {title} {On
  coherent-vortex identification in turbulence},}\ }\href {\doibase
  10.1088/1468-5248/1/1/011} {\bibfield  {journal} {\bibinfo  {journal}
  {Journal of Turbulence}\ }\textbf {\bibinfo {volume} {1}} (\bibinfo {year}
  {2000}),\ 10.1088/1468-5248/1/1/011}\BibitemShut {NoStop}%
\bibitem [{\citenamefont {Anderson}, \citenamefont {DuPont},\ and\
  \citenamefont {DebRoy}(2010)}]{Anderson2010Origin}%
  \BibitemOpen
  \bibfield  {author} {\bibinfo {author} {\bibfnamefont {T.~D.}\ \bibnamefont
  {Anderson}}, \bibinfo {author} {\bibfnamefont {J.~N.}\ \bibnamefont
  {DuPont}}, \ and\ \bibinfo {author} {\bibfnamefont {T.}~\bibnamefont
  {DebRoy}},\ }\bibfield  {title} {\enquote {\bibinfo {title} {{Origin of stray
  grain formation in single-crystal superalloy weld pools from heat transfer
  and fluid flow modeling}},}\ }\href {\doibase 10.1016/j.actamat.2009.10.051}
  {\bibfield  {journal} {\bibinfo  {journal} {Acta Materialia}\ }\textbf
  {\bibinfo {volume} {58}},\ \bibinfo {pages} {1441--1454} (\bibinfo {year}
  {2010})}\BibitemShut {NoStop}%
\bibitem [{\citenamefont {De}\ and\ \citenamefont
  {DebRoy}(2003)}]{De2003Probing}%
  \BibitemOpen
  \bibfield  {author} {\bibinfo {author} {\bibfnamefont {A.}~\bibnamefont
  {De}}\ and\ \bibinfo {author} {\bibfnamefont {T.}~\bibnamefont {DebRoy}},\
  }\bibfield  {title} {\enquote {\bibinfo {title} {{Probing unknown welding
  parameters from convective heat transfer calculation and multivariable
  optimization}},}\ }\href {\doibase 10.1088/0022-3727/37/1/023} {\bibfield
  {journal} {\bibinfo  {journal} {Journal of Physics D: Applied Physics}\
  }\textbf {\bibinfo {volume} {37}},\ \bibinfo {pages} {140+} (\bibinfo {year}
  {2003})}\BibitemShut {NoStop}%
\bibitem [{\citenamefont {De}\ and\ \citenamefont
  {DebRoy}(2004)}]{De2004Smart}%
  \BibitemOpen
  \bibfield  {author} {\bibinfo {author} {\bibfnamefont {A.}~\bibnamefont
  {De}}\ and\ \bibinfo {author} {\bibfnamefont {T.}~\bibnamefont {DebRoy}},\
  }\bibfield  {title} {\enquote {\bibinfo {title} {{A smart model to estimate
  effective thermal conductivity and viscosity in the weld pool}},}\ }\href
  {\doibase 10.1063/1.1695593} {\bibfield  {journal} {\bibinfo  {journal}
  {Journal of Applied Physics}\ }\textbf {\bibinfo {volume} {95}},\ \bibinfo
  {pages} {5230--5240} (\bibinfo {year} {2004})}\BibitemShut {NoStop}%
\bibitem [{\citenamefont {De}\ and\ \citenamefont
  {DebRoy}(2006)}]{De2006Improving}%
  \BibitemOpen
  \bibfield  {author} {\bibinfo {author} {\bibfnamefont {A.}~\bibnamefont
  {De}}\ and\ \bibinfo {author} {\bibfnamefont {T.}~\bibnamefont {DebRoy}},\
  }\bibfield  {title} {\enquote {\bibinfo {title} {{Improving reliability of
  heat and fluid flow calculation during conduction mode laser spot welding by
  multivariable optimisation}},}\ }\href {\doibase 10.1179/174329306x84346}
  {\bibfield  {journal} {\bibinfo  {journal} {Science and Technology of Welding
  and Joining}\ }\textbf {\bibinfo {volume} {11}},\ \bibinfo {pages} {143--153}
  (\bibinfo {year} {2006})}\BibitemShut {NoStop}%
\bibitem [{\citenamefont {Pavlyk}\ and\ \citenamefont
  {Dilthey}(2001)}]{Pavlyk2001Numerical}%
  \BibitemOpen
  \bibfield  {author} {\bibinfo {author} {\bibfnamefont {V.}~\bibnamefont
  {Pavlyk}}\ and\ \bibinfo {author} {\bibfnamefont {U.}~\bibnamefont
  {Dilthey}},\ }\bibfield  {title} {\enquote {\bibinfo {title} {{A numerical
  and experimental study of fluid flow and heat transfer in stationary GTA weld
  pools}},}\ }in\ \href@noop {} {\emph {\bibinfo {booktitle} {Mathematical
  Modelling of Weld Phenomena 5}}},\ \bibinfo {series and number} {Materials
  Modelling Series},\ \bibinfo {editor} {edited by\ \bibinfo {editor}
  {\bibfnamefont {H.}~\bibnamefont {Cerjak}}\ and\ \bibinfo {editor}
  {\bibfnamefont {H.~K. D.~H.}\ \bibnamefont {Bhadeshia}}}\ (\bibinfo
  {publisher} {IOM Communications Ltd},\ \bibinfo {year} {2001})\ pp.\ \bibinfo
  {pages} {135--163}\BibitemShut {NoStop}%
\bibitem [{\citenamefont {Tan}, \citenamefont {Bailey},\ and\ \citenamefont
  {Shin}(2012)}]{Tan2012Numerical}%
  \BibitemOpen
  \bibfield  {author} {\bibinfo {author} {\bibfnamefont {W.}~\bibnamefont
  {Tan}}, \bibinfo {author} {\bibfnamefont {N.~S.}\ \bibnamefont {Bailey}}, \
  and\ \bibinfo {author} {\bibfnamefont {Y.~C.}\ \bibnamefont {Shin}},\
  }\bibfield  {title} {\enquote {\bibinfo {title} {{Numerical Modeling of
  Transport Phenomena and Dendritic Growth in Laser Spot Conduction Welding of
  304 Stainless Steel}},}\ }\href {\doibase 10.1115/1.4007101} {\bibfield
  {journal} {\bibinfo  {journal} {Journal of Manufacturing Science and
  Engineering}\ }\textbf {\bibinfo {volume} {134}},\ \bibinfo {pages} {041010+}
  (\bibinfo {year} {2012})}\BibitemShut {NoStop}%
\end{thebibliography}%

\end{document}